\documentclass[reprint,amsmath,amssymb,aps]{revtex4-2}
\usepackage[utf8]{inputenc}
\usepackage{multirow}
\usepackage{amsmath}
\usepackage{amsthm}
\usepackage{amssymb}
\usepackage{graphicx}

\makeatletter

\providecommand{\tabularnewline}{\\}

\usepackage{dcolumn}
\usepackage{bm}
\usepackage{braket}
\usepackage{amsthm}

\makeatother

\begin{document}
\title{Orbital Energies Are Chemical Potentials in Ground-State
Density Functional Theory and Excited-State $\Delta$SCF Theory}
\author{Weitao Yang}
\affiliation{Department of Chemistry and Department of Physics, Duke University,
Durham, North Carolina 27708}
\author{Yichen Fan}
\affiliation{Department of Chemistry, Duke University, Durham, North Carolina 27708}
\begin{abstract}
We prove the general chemical potential theorem: the noninteracting
one-electron orbital energies in DFT ground states and $\Delta$SCF
excited states are corresponding chemical potentials of electron addition
or removal, from an $N$-particle ground or excited state to an $(N\pm1)$-particle
ground or excited state. This greatly extends the previous ground state results.
Combining with the recently developed exact linear conditions for
fractional charges in excited states, where the slopes of the linear
lines are defined as the excited-state chemical potentials, our result
establish the physical meaning of orbital energies as
approximation to the corresponding excited-state ionization potentials
and electron affinities, for both ground and excited states of a molecule
or a bulk system. To examine the quality of this approximation we
demonstrate numerically significant delocalization error in commonly
used functionals and excellent agreement in functionals correcting
the delocalization error. 
\end{abstract}
\maketitle
Density functional theory (DFT) was formulated as a ground-state theory
and has since become the primary tool for electronic structure calculations \cite{Hohenberg64B864,Kohn65A1133,Parr89,Dreizler90,Teale2228700}. The accuracy of DFT computations hinges critically on the employed
density functional approximations (DFAs). Developing improved functionals
necessitates a deep understanding of the exact conditions they must
fulfill \cite{Perdew963865a,Burke12150901,Cohen12289,Becke1418A301,Sun15036402a,Wang202294,Mardirossian16214110}.

The extension to fractional charges and the exact linear conditions
for the energy were developed by Perdew, Parr, Levy and Baldus (PPLB)
based on grand canonical ensembles \cite{Perdew821691a,Zhang00346}.
This led to the conclusion that the exact chemical potentials, as
the derivative of total energy with respect to electron number has
two limits for an integer particle system: it is equal to the negative
of ionization potential (IP) on the electron removal side, and equal
to the negative of electron affinity (EA) on the electron addition
side \cite{Perdew821691a}. The PPLB conditions were subsequently
derived based on the exact properties of quantum mechanical degeneracy
and size consistency \cite{Yang005172}, and extended to the combination
of fractional charges and spins \cite{Cohen08121104,Mori-Sanchez09066403a}.
The PPLB conditions play an important role in understanding the systematic
errors in commonly used DFAs and in developing their corrections.

A major systematic error in DFAs is the delocalization error (DE)
\cite{Mori-Sanchez08146401}, which leads to the underestimation
of band gaps, reaction barriers, binding energies of charge-transfer
complexes, and dissociation energies, as well as the overestimation
of polymer polarizabilities \cite{Cohen08792,Cohen12289,Bryenton22e1631a}.
The DE has a size-dependent manifestation: for systems small both
in number of atoms and in physical extent, DE appears as the convex
deviation from the PPLB linearity conditions; but for large systems
and bulk systems, DE leads to the incorrect total energy differences
with the ($N\pm1)$ charged states and consequently the underestimation
of band gaps for bulk systems \cite{Mori-Sanchez08146401,Mei22acs.jctc.1c01058}.
To address these challenges, numerous correction methods have been
proposed \cite{Perdew815048,Pederson14121103a, Savin96327,Iikura013540,Song09144108,Wagle21094302c,Baer1085,Wing21,Su149201b,Su162285a, Li244876,Anisimov05075125,Ma1624924,Dabo10115121,Colonna22,Cohen07191109,Zheng11,Li15053001,Li18203,Su201528,Mei22acs.jctc.1c01058}.

In the ground state KS theory, while the electron density is the basic
but implicit variable, the noninteracting reference system provides
the explicit variable for defining the kinetic and exchange-correlation
energy functionals, and the electron density or the density matrix.
The noninteracting reference system is associated with a noninteracting
Hamiltonian, the eigenstates of which are the one-electron Kohn-Sham
(KS) orbitals with corresponding orbital energies. The Janak theorem
established the equality of orbital energies with the derivatives
of total energy with respect to the orbital occupation numbers \cite{Janak787165}.
This is very important, but not directly connecting orbital energies
to experimental observables. The physical meaning of the orbital energies
is most interesting and has been explored over a long time from the
conditions of the exact functional and exact asymptotic density behavior
\cite{Morrell75549a,Perdew821691a,Perdew9716021b,Tozer002117b}, the connection
with Dyson orbitals and ionization energies \cite{Chong021760,Gritsenko031937a},
the observed approximation to optical excitation energies \cite{Savin98391,vanMeer144432},
and the equality of frontier orbital energies with the chemical potentials
(thus approximation to the ionization energy and electron affinity)
\cite{Cohen08115123,Cohen12289,Yang12204111}.

In ground state DFT, two relationships connecting orbital energies to experimental observables
have been established rigorously: (1) \emph{The ionization potential
theorem} \cite{Perdew821691a}: For the \emph{exact} and \emph{local} Kohn-Sham potential, the  KS highest occupied molecular orbital (HOMO) energy is
negative of the first ionization energy  \cite{Perdew821691a,Perdew9716021b}.
(2) \emph{The ground state chemical potential theorem} \cite{Cohen08115123}:
In any ground-state DFA calculation with an $E_{xc}$ functional that
is a continuous functional of density or the KS density matrix, the
energy of the HOMO is the chemical potential of the electron removal and the energy of the lowest unoccupied molecular orbital (LUMO) is the chemical potential of electron addition \cite{Cohen08115123,Cohen12289,Yang12204111}. The ground state chemical potential theorem justifies using
the frontier orbital energies to approximate experimental IP and EA in molecules and valence and conduction band edge energies and hence band gaps
in bulk systems, as the exact ground state chemical potentials are -IP and -EA
from the PPLB condition \cite{Cohen08115123,Cohen12289,Yang12204111}.
The result is applicable both for the KS calculations with a local
potential when the $E_{xc}$ is given as a functional of the density,
and for the generalized KS calculation with a nonlocal potential when
the $E_{xc}$ is given as a functional of the density matrix of the
noninteracting reference system. Particularly, the ground state chemical
potential theorem for the first time established the physical meaning
of the LUMO energy in ground state (G)KS calculations, supported with
numerical results \cite{Cohen08115123,Yang12204111}. Further discussion for bulk systems can be found in Ref.\cite{Perdew172801}.

The meaning of the other orbitals has also been much explored. For
the occupied orbitals below HOMO, connections to the higher ionization
energies has been discussed and observed in some DFA calculations
with good agreement \cite{Chong021760,Gritsenko031937a}. For the
unoccupied orbitals above LUMO, good agreement to the optical excitation
energies was observed in atomic calculations with accurate KS local
potentials \cite{Savin98391}, which was further explored \cite{vanMeer144432}.

Consistent with the ground-state chemical potential theorem, recent
functional developments in reducing DE lead to results in excellent
agreement of frontier orbital energies with the IP and EA in molecules,
and band gaps for bulk systems, similar or better than the GW Green's
function results \cite{Cohen08115123,Zheng11,Li15053001,Li18203,Su2020,  Savin96327,Song09144108,Hirao213489,Ma1624924,Dabo10115121,Borghi14075135,Colonna182549,Nguyen18021051,Colonna191905,Colonna225435,Wing21,Williams24,Yu24}.
Remarkably, similar agreements also have been reported for the energies
of orbitals below HOMO with the negative of corresponding higher IPs
\cite{Mei19666}, even for core orbitals \cite{Yu24}, supporting the
interpretation for all orbital energies as quasiparticle energies
\cite{Mei19666}. This quasiparticle energy interpretation
is consistent with the IP and EA connection to the HOMO and LUMO energies
and would present a unified physical meaning for the \emph{entire}
orbital spectrum, much like the Koopmans' theorem for Hartree-Fock
theory, under the frozen orbital assumption \cite{Koopmans34104}. Furthermore, two groups have independently
developed the method for calculating optical excitation energies of
an $N$-particle system based on the ground state orbital energies
of an $N-1$or $N+1$systems, using quasipaticle energies as approximated
from the corresponding ground state orbital energies \cite{Haiduke18131101,Mei18b,Mei19666}.
This method was called quasiparticle energy DFT (QE-DFT) and has been
shown to describe well valence and Rydberg excitations \cite{Mei19666},
and charge transfer excitations \cite{Mei19144109}, conical intersections
\cite{Mei192538} and excited-state charge transfer coupling \cite{Kuan246126}.

For the quasiparticle energy interpretation of all orbital energies
of a ground state to be true, the ground state DFT functional has
to contain excited-state information. A abundance of accurate numerical
results supports this interpretation \cite{Mei19666,Yu24}, but  lacking a rigorous theoretical foundation. This motivated us to seek for the understanding, leading to the development of theoretical foundation for the $\Delta$SCF excited state approach \cite{Yang24}, the extension of the excited
state theory to fractional charges and the proof of the exact linear
conditions \cite{yang_fractional_2024}, and the present work. The key result of
present work is the establishment of the general chemical potential
theorem on the physical meanings of all orbital energies, both for
ground state (G)KS calculations while encompassing the previous ground
state chemical potential theorem, and for excited state $\Delta$SCF
calculations. Before we proceed, we need to review related developments
in excited state theory.

Indeed, deviating from the ground-state formulation, KS DFT has been
extensively employed for excited-state calculations via the $\Delta$SCF
approach for a long time\cite{Slater1974AdvQuantumChem,SlaterWood1970SlaterTM}.
It has demonstrated significant numerical success in predicting excitation
energies \cite{ZieglerRaukBaerends1977deltaSCF,GavnholtOlsen2008DeltaSCFes,GilbertBesleyGill2008MOM,BesleyGilbertGill2009DeltaSCF,HarbolaSamal2009esHK,KowalczykanVoorhis2011deltaSCF,MaurerReuter2011deltaSCF,MaurerReuter2013deltaSCF,SeiduZiegler2015constrictedDFT,YeWelbornVanVoorhis2017deltaSCFalgorithm,HaitHeadgordon2020deltaSCF,CarterfenkHerbert2020deltaSCFnumerical,LeviJonsson2020deltaSCF,CorzoPribramjonesHratchian2022PMOM,KumarLuber2022deltaSCF,VandaeleLuber2002deltaSCFreview},
albeit lacking a rigorous theoretical foundation \cite{VandaeleLuber2002deltaSCFreview},
until recently \cite{Yang24}.

To describe excited states, the electron density alone is insufficient.
Instead, a set of equivalent variables defining the non-interacting
reference system can be utilized: the excitation number $n_{s}$ and
the local one-electron potential $w_{s}({\bf x})$, the noninteracting
wavefunction $\Phi,$ or the 1-particle density matrix $\gamma_{s}({\bf x},{\bf x}')$
\cite{Yang24}. While the electron density is no longer the fundamental
variable, it remains crucial as the physical system's density equals
that of the non-interacting reference system for both ground states
\cite{Hohenberg64B864,Kohn65A1133,Parr89,Dreizler90} and excited
states \cite{Yang24}. Ground and excited state energies and densities
are obtained from the minimum and stationary solutions of the \emph{same}
universal functional. 

Using $\gamma_{s}$ as the basic variable for describing excited states
in the density matrix functional ($\gamma_{s}$FT) \cite{Yang24},
the excited-state theory has been recently extended to fractional
charges for excited states \cite{yang_fractional_2024}, following the previous
approach for ground states \cite{Yang005172}. Consider two many-electron
systems with the same external potential $v(\mathbf{r})$, one with
$N$ electron in the $n$th excited state, and the other with $N+1$
electrons in the $m$th excited state. The energies of these two states
are $E_{v}^{n}(N)$ and $E_{v}^{m}(N+1)$ respectively. Within the
$\Delta$SCF theory, as formulated recently, the corresponding non-interacting
reference systems have the first-order density matrices $\gamma_{s}^{n_{s}}(N)$,
 and $\gamma_{s}^{m_{s}}(N+1)$ respectively, where $n_{s}$ and
$m_{s}$ are the corresponding excitation numbers of the non-interacting
reference systems \cite{Yang24}. Parallel to the ground state fractional
charges in terms of electron densities, the fractional charge system
for excited states is described by the density matrix, for $0\leq\delta\leq1$,

\begin{equation}
\gamma_{s}=(1-\delta)\gamma_{s}^{n_{s}}(N)+\delta\gamma_{s}^{m_{s}}(N+1)\label{eq:gammaFractional}
\end{equation}
The following linear conditions has been proved for the exact energy
functional \cite{yang_fractional_2024}: 
\begin{align}
 & E_{v}[(1-\delta)\gamma_{s}^{n_{s}}(N)+\delta\gamma_{s}^{m_{s}}(N+1)]\nonumber \\
= & (1-\delta)E_{v}^{n}(N)+\delta E_{v}^{m}(N+1).\label{GeneralLinearity}
\end{align}
Note that the excited state excitation levels $n$ and $m$ correspond
to the excitation levels $n_{s}$ and $m_{s}$ of the noninteracting
reference system -- they do not need to be the same \cite{Yang24}.
This result agrees with PPLB linear conditions in the special case
of ground states: $n_{s}=m_{s}=n=m=0.$ It extends the PPLB linear
conditions in two key ways: the basic variables for the fractional
charge systems are now $\gamma_{s},$ the 1-particle density matrices
of the noninteracting reference systems, and the $N$- and $\left(N+1\right)$-electron
states are all states, ground and excited.

For ground states, the PPLB linear conditions set the physical meaning
for ground-state chemical potentials \cite{Parr783801}, $\mu=\left(\frac{\partial E_{v}(\mathit{\mathcal{N}})}{\partial\mathit{\mathcal{N}}}\right)_{v}$,
the slopes of $E_{v}(\mathit{\mathcal{N}})$ ($\mathit{\mathcal{N}}$denotes
fractional electron numbers): $\mu^{-}=-IP$ and $\mu^{+}=-EA$, where
the ionization potential $IP=E_{v}(N-1)-E_{v}(N)$ and the electron
affinity $EA=E_{v}(N)-E_{v}(N+1)$ \cite{Perdew821691a,Parr89}.

For excited states, the slopes of the linear curves in Eq. (\ref{GeneralLinearity})
also convey physical meanings. For a given $n_{s}$th eigenstate with
an integer particle number $N,$ the fractional electron number $\mathit{\mathcal{N}}$
connecting to the $m_{s}$th eigenstate of the $(N+1)-$electron system
(described by $\gamma_{s}^{m_{s}}(N+1)$) is $\mathit{\mathcal{N}}=(1-\delta)N+\delta(N+1)$,
and the energy as a function of $\mathit{\mathcal{N}}$ is $E_{v}^{+}(n_{s},m_{s},\mathit{\mathcal{N}})=E_{v}[(1-\delta)\gamma_{s}^{n_{s}}(N)+\delta\gamma_{s}^{m_{s}}(N+1)],$
the left hand side of Eq. (\ref{GeneralLinearity}). The excited-state
chemical potential $\mu_{n_{s}m_{s}}^{+}$ is the slope of $E_{v}^{+}(n_{s},m_{s},\mathit{\mathcal{N}})$:
For $N\leq\mathcal{N}\leq N+1$,\\

\begin{equation}
\mu_{n_{s}m_{s}}^{+}(\mathcal{N})=\left(\frac{\partial E_{v}^{+}(n_{s},m_{s},\mathit{\mathcal{N}})}{\partial\mathit{\mathcal{N}}}\right)_{v}=E_{v}^{m}(N+1)-E_{v}^{n}(N).\label{eq:mu_plus}
\end{equation}

Similarly on the electron removal side, the fractional electron number
$\mathit{\mathcal{N}}$ connecting to the $l_{s}$th eigenstate of
the $(N-1)-$electron system (described by $\gamma_{s}^{l_{s}}(N-1)$)
is $\mathit{\mathcal{N}}=(1-\delta)N+\delta(N-1)$, and the energy
as a function of $\mathit{\mathcal{N}}$ is $E_{v}^{-}(n_{s},l_{s},\mathit{\mathcal{N}})=E_{v}[(1-\delta)\gamma_{s}^{n_{s}}(N)+\delta\gamma_{s}^{l_{s}}(N-1)].$
The excited-state chemical potential $\mu_{n_{s}m_{s}}^{-}$ is the
slope of $E_{v}^{-}(n_{s},l_{s},\mathit{\mathcal{N}})$: For $N-1\leq\mathcal{N}\leq N$,

\begin{equation}
\mu_{n_{s}l_{s}}^{-}(\mathcal{N})=\left(\frac{\partial E_{v}^{-}(n_{s},l_{s},\mathit{\mathcal{N}})}{\partial\mathit{\mathcal{N}}}\right)_{v}=E_{v}^{n}(N)-E_{v}^{l}(N-1).\label{eq:mu_minus}
\end{equation}
There is an symmetry: $\mu_{n_{s}m_{s}}^{+}(N)$=$\mu_{m_{s}n_{s}}^{-}(N+1)$
\cite{yang_fractional_2024}. For ground states, $\mu_{00}^{+}$ and $\mu_{00}^{-}$
agree with the chemical potentials, $\mu^{+}$ and $\mu^{-}$, from
the PPLB condition \cite{Perdew821691a,Parr89}. The excited-state
chemical potentials are thus the negative of IP
associated with an excited state with one electron removed as in Eq.
(\ref{eq:mu_minus}), or the negative of EA associated
with an excited state with one added electron as in Eq.(\ref{eq:mu_plus}).

Now we describe present work on the physical meaning of \textit{all} orbital
energies in $\Delta$SCF excited state theory for an $N$-particle
ground or excited state. Consider a $\Delta$SCF calculation with
a given DFA $E_{xc}[\gamma_{s}({\bf x},{\bf x}')]$ for an excited
state with fractional charge of Eq. (\ref{eq:gammaFractional}), using
fractional occupations of $0\leq f_{p}\leq1,$ 
\begin{equation}
\gamma_{s}({\bf x},{\bf x}')=\sum_{p}f_{p}\phi_{p}(\mathbf{x})\phi_{p}^{*}({\bf x}'),\label{eq:Janak_gamma}
\end{equation}
which is consistent with how fractional charge calculations are carried
out in the ground-state (G)KS calculations (\cite{SlaterWood1970SlaterTM,Mori-Sanchez06201102,Ruzsinszky06194112}.
The total energy functional is 
\begin{equation}
E_{v}\left[\gamma_{s}\right]=T_{s}\left[\gamma_{s}\right]+J\left[\rho\right]+E_{xc}[\gamma_{s}]+\int d\mathbf{r}v(\mathbf{x})\rho(\mathbf{x}),\label{eq:TotalE}
\end{equation}
where $T_{s}\left[\gamma_{s}\right]=\textrm{Tr}(t\gamma_{s})$ and
$J\left[\rho\right]$ is the classical Coulomb interaction energy.
The one-electron Hamiltonian for the nonintercating reference system
is 
\begin{equation}
h_{\text{eff}}({\bf x},{\bf x}')=\frac{\delta E_{v}[\gamma_{s}]}{\delta\gamma_{s}(\mathbf{r}',\mathbf{r})}=t+v_{\text{eff}}(\mathbf{x},\mathbf{x}'),\label{eq:h_eff}
\end{equation}
where in general one has an effective nonlocal potential $v_{\text{eff}}(\mathbf{x},\mathbf{x}')=\left(v(\mathbf{x})+v_{J}(\mathbf{r})\right)\delta(\mathbf{x}-\mathbf{x}')+v_{xc}(\mathbf{x},\mathbf{x}'),$with
$v_{xc}(\mathbf{r},\mathbf{r}')=\frac{\delta E_{xc}[\gamma_{s}]}{\delta\gamma_{s}(\mathbf{r}',\mathbf{r})}$.
When one uses a density functional approximation $E_{xc}[\rho]$,
then $v_{xc}(\mathbf{x},\mathbf{x}')=\frac{\delta E_{xc}[\rho]}{\delta\rho(\mathbf{r})}\delta(\mathbf{x},\mathbf{x}')=v_{xc}(\mathbf{x})\delta(\mathbf{x},\mathbf{x}')$.

Consider the derivative of the total energy with respect to the orbital
occupation $f_{q}$,

\begin{align}
\frac{\partial E_{v}\left[\gamma_{s}\right]}{\partial f_{q}} & =\int d{\bf x}d{\bf x}'\frac{\delta E_{v}[\gamma_{s}]}{\delta\gamma_{s}({\bf x},{\bf x}')}\left(\frac{\partial\gamma_{s}({\bf x},{\bf x}')}{\partial f_{q}}\right)\nonumber \\
 & =\left\langle \phi_{q}\left|h_{\text{eff}}\right|\phi_{q}\right\rangle =\varepsilon_{q}\label{eq:JanakExcitedState}
\end{align}
where we used   $h_{\text{eff}}\left|\phi_{q}\right\rangle =\varepsilon_{q}\left|\phi_{q}\right\rangle $
which is satisfied upon SCF convergence and the fact that orbital derivative
terms sum to zero: $\left\langle \phi_{p}\left|h_{\text{eff}}\right|\frac{\partial\phi_{p}}{\partial f_{q}}\right\rangle +\left\langle \frac{\partial\phi_{p}}{\partial f_{q}}\left|h_{\text{eff}}\right|\phi_{p}\right\rangle =\varepsilon_{p}\frac{\partial}{\partial f_{q}}\left\langle \phi_{p}\right|\left.\phi_{p}\right\rangle =0.$
Eq. (\ref{eq:JanakExcitedState}) is a direct generalization of Janak's
result to $\Delta$SCF excited state theory for fractional particle
numbers, and with extension for nonlocal potential GKS calculations (which was previously developed 
for ground states \cite{Cohen08792,Mei22acs.jctc.1c01058}). It is thus
valid for all orbitals with any fractional occupation, with local
or nonlocal one-electron potential and for ground-state and excited
state $\Delta$SCF calculations.

For a physical system with integer particle numbers in its $n$th
excited state described by the $n_{s}$th excited state of of a noninteracting
system in $\Delta$SCF approach, we consider the removal of an infinitesimal
charge $\delta f_{i}$ from an occupied orbital $i$. This moves our
system from the original $N$-particle excited state $n_{s}$ towards
the $(N-1)$-particle excited state $l_{s}$, the associated chemical
potential is

\begin{align}
\mu_{n_{s}l_{s}}^{-}(N) & =\left.\left(\frac{\partial E_{v}^{-}(n_{s},l_{s},\mathit{\mathcal{N}})}{\partial\mathit{\mathcal{N}}}\right)_{v}\right|_{f_{i}=1}\nonumber \\
 & =\left.\frac{\partial E_{v}\left[\gamma_{s}\right]}{\partial f_{i}}\right|_{f_{i}=1}=\varepsilon_{i}(N),\label{eq:mu-}
\end{align}
which is just the orbital energy of the $i$th occupied state. Similarly
for an unoccupied orbital $a$ ,

\begin{align}
\mu_{n_{s}m_{s}}^{+}(N) & =\left.\left(\frac{\partial E_{v}^{+}(n_{s},m_{s},\mathit{\mathcal{N}})}{\partial\mathit{\mathcal{N}}}\right)_{v}\right|_{f_{a}=0}\nonumber \\
 & =\left.\frac{\partial E_{v}\left[\gamma_{s}\right]}{\partial f_{a}}\right|_{f_{a}=0}=\varepsilon_{a}(N).\label{eq:mu+}
\end{align}
\begin{figure}[ht!]
\begin{centering}
\includegraphics[width=3.4in]{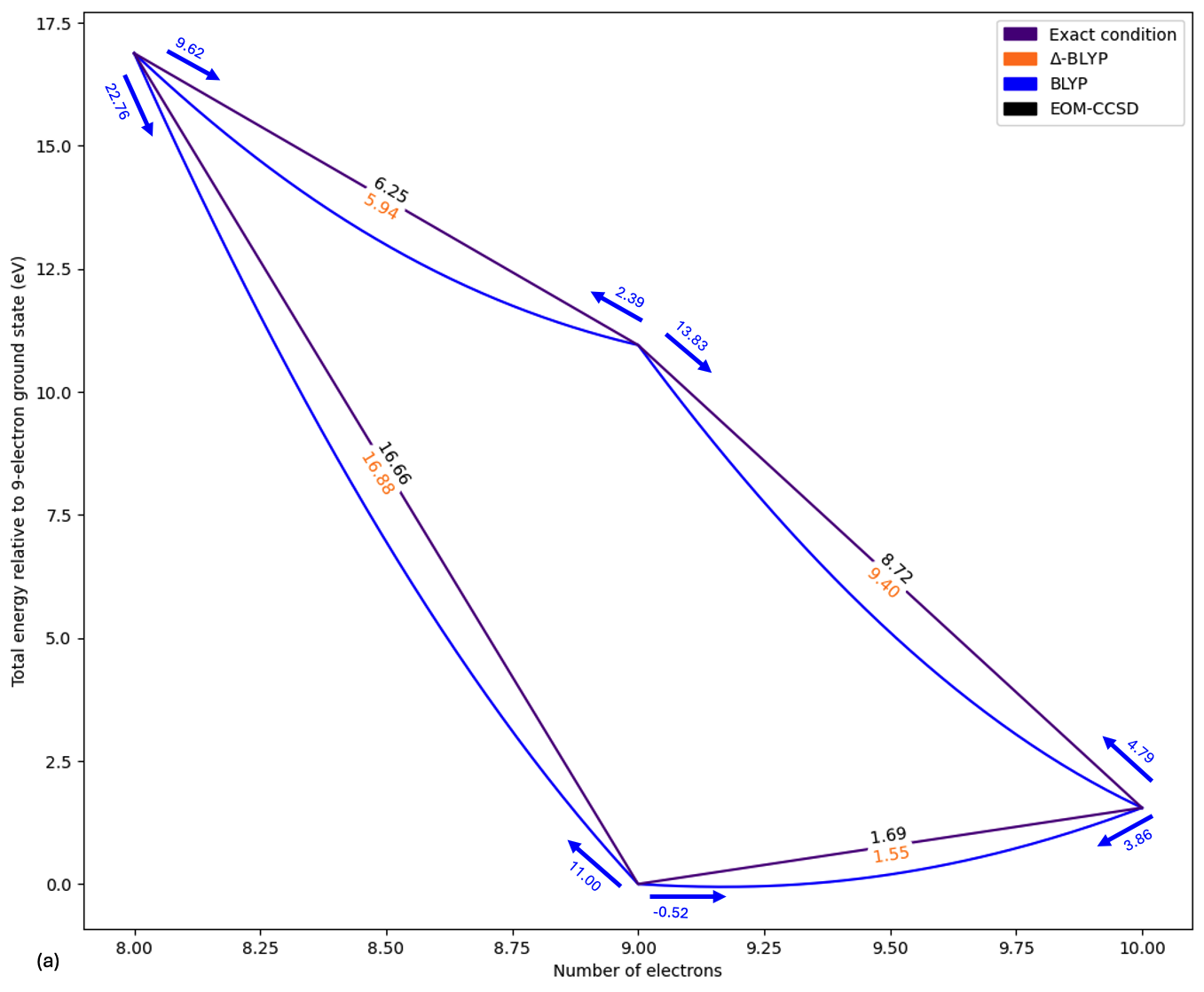} 
\par\end{centering}
\begin{centering}
\includegraphics[width=3.4in]{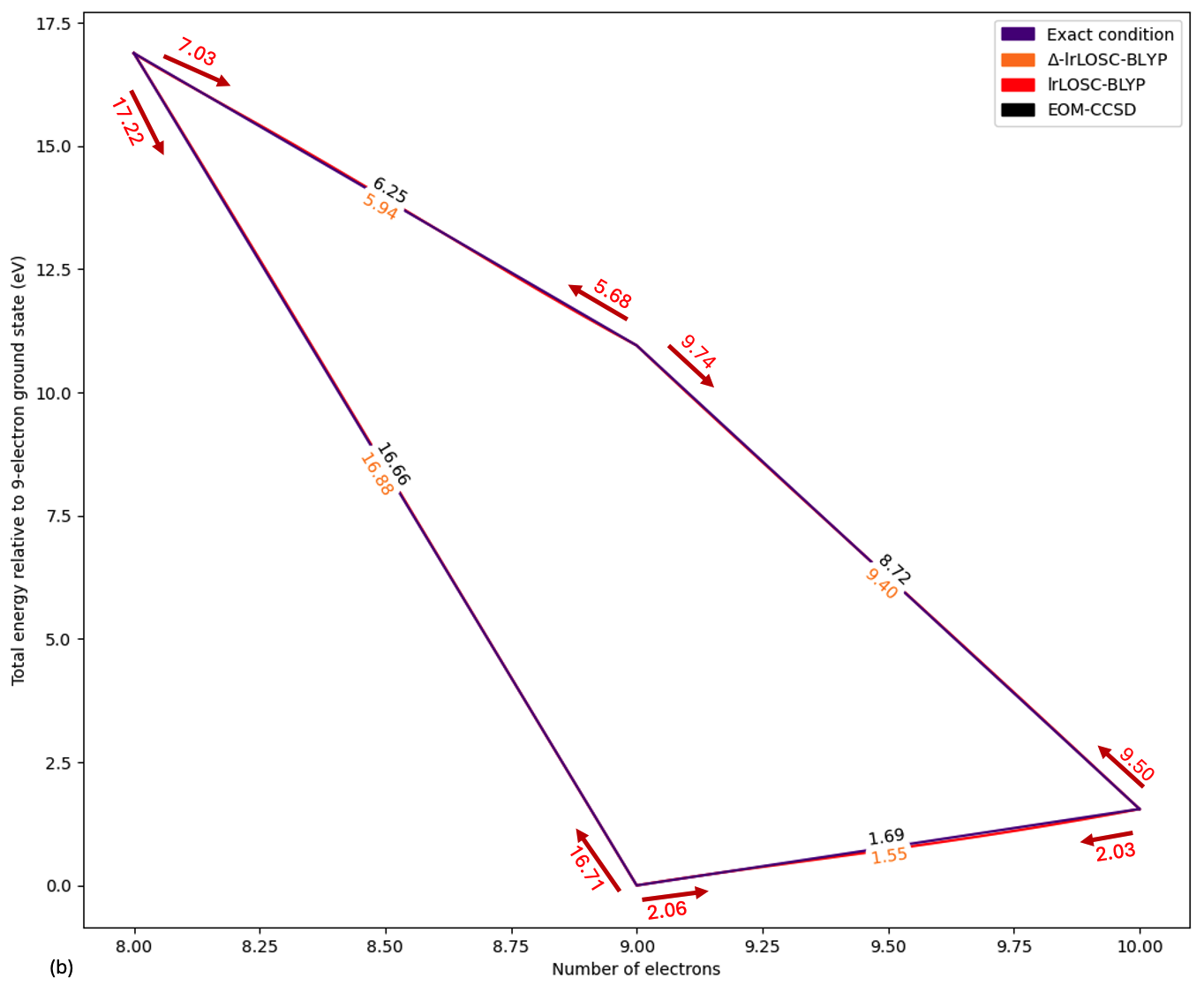} 
\par\end{centering}
\centering{}\caption{Excited-state energies for fractional charges and excited-state chemical
potentials calculated from orbital energies in transitions between
three systems in four states, $\text{OH}^{+}$ (N=8, excited state
$^{3}\Pi$ ), $\text{OH}$ (N=9, ground state $^{2}\Pi$ and excited
state $^{4}\Pi$), and $\text{OH}^{-}$ (N=10, $^{3}\Pi$ excited
state). Orbital energies for BLYP\cite{Becke1988, Lee1988} (a, blue) and lrLOSC-BLYP\cite{Li18203,Su201528,Mei2021, Yu24, Williams24} (b, red)
are shown with numbers and arrows indicating the + or the - sign for
the corresponding chemical potentials. Additionally, $\Delta$-BLYP
(a,orange), $\Delta$-lrLOSC-BLYP (b, orange)and the reference EOM-CCSD
results (black) are shown as inline labels.}
\end{figure}

\null

The initial and the final excitation levels, $n_{s}$, $l_{s}$ and
$m_{s}$, are uniquely determined by the three sets of integer orbital
occupations corresponding to the initial $N$electron state and the
final $(N\mp1)$ states, together with the spectrum of the corresponding  three
distinctive noninteracting Hamiltonians. Note that within $\Delta$SCF,
each excited state of a physical system has its own inoninteracting
reference system and a unique  $h_{\text{eff}}$. While the orbital
index $i$ for electron removal and $a$ for electron addition do
not themselves explicitly lead to the excitation levels $l_{s}$ and
$m_{s}$ for the final $(N\mp1)$ states, they do so in combination
with the spectrum of each $h_{\text{eff}}$. The symmetry relation
in chemical potentials $\mu_{n_{s}m_{s}}^{+}(N)$=$\mu_{m_{s}n_{s}}^{-}(N+1)$
for the exact theory will also be reflected in the corresponding symmetry
of orbital energies $\varepsilon_{a}(N)$ and $\varepsilon_{i}(N+1)$.

Eqs. (\ref{eq:mu-}-\ref{eq:mu+}), exact for any continuous (approximate) DFAs, are the key results of the general
chemical potential theorem: It equates an occupied/unocxcupied orbital energy to the corresponding excited-state chemical potential of electron
removal/addition. It greatly
extends the previous results that are limited to all ground-state
quantities \cite{Cohen08115123}. With the exact conditions for excited-state
chemical potentials, Eqs. (\ref{eq:mu_plus}-\ref{eq:mu_minus}),
the orbital energies are thus the DFA approximation to the corresponding
excited-state IP, or excited-state quasiparticle energy, 
\begin{equation}
\varepsilon_{i}(N)=E_{v}^{n}(N)-E_{v}^{l}(N-1).\label{eq:exQuasiH}
\end{equation}
and excited-state EA, or excited-state quasiparticle energy, 
\begin{equation}
\varepsilon_{a}(N)=E_{v}^{m}(N+1)-E_{v}^{n}(N).\label{eq:exQuasiP}
\end{equation}
When starting from a ground state, $n=n_{s}=0$, we have $\varepsilon_{i}(N)=E_{v}^{0}(N)-E_{v}^{l}(N-1),$ and $\varepsilon_{a}(N)=E_{v}^{m}(N+1)-E_{v}^{0}(N).$
This provides the theoretical foundation for the quasiparticle energy interpretation of all orbital energies
of ground-state calculations and thus for using orbital energies
in  predicting photoemission and inverse photoemission spectra of
ground state systems \cite{Mei19666}. It also justifies the QE-DFT approach  \cite{Haiduke18131101,Mei18b,Mei19666} to calculating
optical excitation energies of an $N$-particle system from $(N\pm1)$-particle ground-state orbital energies. 

Before this work, no interpretation of the orbital energies had
 been reported for excited state $\Delta$SCF calculations. Eqs. (\ref{eq:exQuasiH}-\ref{eq:exQuasiP}), going beyond QE-DFT,   provides  a new approach, which we call excited state quasiparticle energy from DFT (exQE-DFT), to predicting excitation energies of the $(N\pm1)$-particle excited states, from the orbital energies of an  $N$-particle excited state  $\Delta$SCF calculation,  allowing efficient computational approach to broader excitations.   

For optimized effective potential (OEP) calculations, the orbital
energies are in general not equal to the chemical potentials for ground
states \cite{Cohen08792} and similarly for excited states (See SI).
The quality of the approximation of orbital energies to the corresponding
excited-state IPs/EAs depends on the exchange-correlation energy
functional approximations used, as expected from previous results for ground
state quantities \cite{Cohen08792}.

\begin{table}[ht!]
\centering %
\begin{tabular}{cc|cccc}
\hline 
Molecular  & Chemical  &  &  &  & EOM- \tabularnewline
Orbital  & Potential  & $\varepsilon_{\text{PBE}}$  & $\varepsilon_{\text{lrLOSC-PBE}}$  & $\Delta$-PBE  & CCSD \tabularnewline
\hline 
{$(\sigma_{\beta}^{*})^{0}$}  & {$\mu_{^{2}\Pi,^{1}\Pi}^{+}$}(13)  & 11.382  & 6.432  & \multirow{2}{*}{6.548} & \multirow{2}{*}{6.855}\tabularnewline
{$(\sigma_{\beta}^{*})^{1}$}  & {$\mu_{^{1}\Pi,^{2}\Pi}^{-}$}(14)  & 1.934  & 6.269  &  & \tabularnewline
$(\sigma_{\alpha}^{*})^{0}$  & $\mu_{^{2}\Pi,^{3}\Pi}^{+}$(13)  & 13.492  & 8.377  & \multirow{2}{*}{8.410} & \multirow{2}{*}{8.656}\tabularnewline
$(\sigma_{\alpha}^{*})^{1}$  & $\mu_{^{3}\Pi,^{2}\Pi}^{-}$(14)  & 3.506  & 8.150  &  & \tabularnewline
$(\sigma_{\beta}^{*})^{0}$  & $\mu_{^{2}\Sigma^{+},^{1}\Sigma^{+}}^{+}$(13)  & 18.364  & 14.287  & \multirow{2}{*}{13.853} & \multirow{2}{*}{14.165}\tabularnewline
$(\sigma_{\beta}^{*})^{1}$  & $\mu_{^{1}\Sigma^{+},^{2}\Sigma^{+}}^{-}$(14)  & 9.039  & 13.882  &  & \tabularnewline
$(\pi_{\beta})^{0}$  & $\mu_{^{2}\Pi,^{1}\Sigma^{+}}^{+}$(13)  & 22.624  & 17.315  & \multirow{2}{*}{17.223} & \multirow{2}{*}{17.039}\tabularnewline
$(\pi_{\beta})^{1}$  & $\mu_{^{1}\Sigma^{+},^{2}\Pi}^{-}$(14)  & 11.835  & 17.065  &  & \tabularnewline
\end{tabular}\caption{Excited state chemical potentials
for carbon monoxide (CO) in eV. The first column
shows the relevant molecular orbital with its occupation number in
super script. The chemical potentials are expressed with molecular
term symbols to label the excited states involved, with the number
of electrons in parenthesis to indicate the particle number in the
$\Delta$SCF calculation. The table includes chemical potentials between
the ground state ($^{1}\Sigma^{+}$) and the excited states ($^{1}\Pi,^{3}\Pi$)
of $\text{CO}$ and the ground state ($^{2}\Sigma^{+}$) and the excited
state ($^{2}\Pi$) of $\text{CO}^{+}$. The spin purification \cite{ZieglerRaukBaerends1977deltaSCF,KowalczykanVoorhis2011deltaSCF}
has been applied to $\mu_{^{2}\Pi,^{1}\Pi}^{+}$ and $\mu_{^{1}\Pi,^{2}\Pi}^{-}$
in $\Delta$PBE\cite{Perdew963865a} and orbital energy calculations. PBE has good agreement with EOM-CCSD for chemical potentials from $\Delta$SCF total energy differences but significant exDE for chemical potentials from orbital energies.  lrLOSC-PBE \cite{Li18203,Su201528,Mei2021,Yu24,Williams24} gives essentially the same good total energies as PBE (not listed), but excellent chemical potentials from orbital energies. lrLOSC-PBE orbital energies also approximately satisfy the exact symmetry, eg,
$\mu_{^{2}\Pi,^{1}\Pi}^{+}(13)=6.432\approx\mu_{^{1}\Pi,^{2}\Pi}^{-}(14)=6.269$.
}
\end{table}
While Eqs. (\ref{eq:mu-}-\ref{eq:mu+}) are exact for any continuous (approximate) DFAs,  how well Eqs. (\ref{eq:exQuasiH}-\ref{eq:exQuasiP}) are satisfied depends on the quality of the DFA.   With commonly used DFAs, we numerically demonstrate   significant excited-state delocalization error (exDE) \cite{yang_fractional_2024}, in Figure 1, Table 1 and SI with many more systems.
For small systems, DFAs have good description of the total excitation energies at integer charges and exDE is reflected in the significant and systematic
convex deviation from exact linear lines for fractional charges proven recently \cite{yang_fractional_2024}, and the underestimation/overestimation of excited state IPs/EAs from orbital energies, similarly to ground states\cite{Cohen08792}. Based on the ground state understanding \cite{Mori-Sanchez08146401}, we  expect the underestimation/overestimation of excited state IPs/EAs from orbital energies will persist as systems get larger and approach the bulk limit: Even though the convex deviation from the linear conditions will decrease and disappear, exDE will manifest as the errors in the $\Delta$SCF excited state total energy differences. With functionals correcting DE (LOSC \cite{Li18203,Su201528,Mei2021,Yu24,Williams24}) we show  excellent agreement of excited state IPs/EAs with orbital
energies from excited-state  $\Delta$SCF calculations.

Our numerical results exemplify the general chemical potential theorem established in present work on the physical meaning of energies of all orbitals, occupied and virtual, for all states, ground and excited. 

We acknowledge support from the National Science Foundation (CHE-2154831)
and the National Institute of Health (R01-GM061870).

\newpage\bibliography{FractionalDeltaSCF_2024Notes,ExcitedStateChemicalPotnetial,ExcitedStateChemicalPotnetial_add}

\begin{thebibliography}{94}%
\makeatletter
\providecommand \@ifxundefined [1]{%
 \@ifx{#1\undefined}
}%
\providecommand \@ifnum [1]{%
 \ifnum #1\expandafter \@firstoftwo
 \else \expandafter \@secondoftwo
 \fi
}%
\providecommand \@ifx [1]{%
 \ifx #1\expandafter \@firstoftwo
 \else \expandafter \@secondoftwo
 \fi
}%
\providecommand \natexlab [1]{#1}%
\providecommand \enquote  [1]{``#1''}%
\providecommand \bibnamefont  [1]{#1}%
\providecommand \bibfnamefont [1]{#1}%
\providecommand \citenamefont [1]{#1}%
\providecommand \href@noop [0]{\@secondoftwo}%
\providecommand \href [0]{\begingroup \@sanitize@url \@href}%
\providecommand \@href[1]{\@@startlink{#1}\@@href}%
\providecommand \@@href[1]{\endgroup#1\@@endlink}%
\providecommand \@sanitize@url [0]{\catcode `\\12\catcode `\$12\catcode `\&12\catcode `\#12\catcode `\^12\catcode `\_12\catcode `\%12\relax}%
\providecommand \@@startlink[1]{}%
\providecommand \@@endlink[0]{}%
\providecommand \url  [0]{\begingroup\@sanitize@url \@url }%
\providecommand \@url [1]{\endgroup\@href {#1}{\urlprefix }}%
\providecommand \urlprefix  [0]{URL }%
\providecommand \Eprint [0]{\href }%
\providecommand \doibase [0]{https://doi.org/}%
\providecommand \selectlanguage [0]{\@gobble}%
\providecommand \bibinfo  [0]{\@secondoftwo}%
\providecommand \bibfield  [0]{\@secondoftwo}%
\providecommand \translation [1]{[#1]}%
\providecommand \BibitemOpen [0]{}%
\providecommand \bibitemStop [0]{}%
\providecommand \bibitemNoStop [0]{.\EOS\space}%
\providecommand \EOS [0]{\spacefactor3000\relax}%
\providecommand \BibitemShut  [1]{\csname bibitem#1\endcsname}%
\let\auto@bib@innerbib\@empty
\bibitem [{\citenamefont {Hohenberg}\ and\ \citenamefont {Kohn}(1964)}]{Hohenberg64B864}%
  \BibitemOpen
  \bibfield  {author} {\bibinfo {author} {\bibfnamefont {P.}~\bibnamefont {Hohenberg}}\ and\ \bibinfo {author} {\bibfnamefont {W.}~\bibnamefont {Kohn}},\ }\bibfield  {title} {\bibinfo {title} {Inhomogeneous {Electron} {Gas}},\ }\href {https://doi.org/10/csx7jx} {\bibfield  {journal} {\bibinfo  {journal} {Physical Review}\ }\textbf {\bibinfo {volume} {136}},\ \bibinfo {pages} {B864} (\bibinfo {year} {1964})}\BibitemShut {NoStop}%
\bibitem [{\citenamefont {Kohn}\ and\ \citenamefont {Sham}(1965)}]{Kohn65A1133}%
  \BibitemOpen
  \bibfield  {author} {\bibinfo {author} {\bibfnamefont {W.}~\bibnamefont {Kohn}}\ and\ \bibinfo {author} {\bibfnamefont {L.~J.}\ \bibnamefont {Sham}},\ }\bibfield  {title} {\bibinfo {title} {Self-{Consistent} {Equations} {Including} {Exchange} and {Correlation} {Effects}},\ }\href {https://doi.org/10/c7zp2p} {\bibfield  {journal} {\bibinfo  {journal} {Physical Review}\ }\textbf {\bibinfo {volume} {140}},\ \bibinfo {pages} {A1133} (\bibinfo {year} {1965})}\BibitemShut {NoStop}%
\bibitem [{\citenamefont {Parr}\ and\ \citenamefont {Yang}(1989)}]{Parr89}%
  \BibitemOpen
  \bibfield  {author} {\bibinfo {author} {\bibfnamefont {R.~G.}\ \bibnamefont {Parr}}\ and\ \bibinfo {author} {\bibfnamefont {W.}~\bibnamefont {Yang}},\ }\href@noop {} {\emph {\bibinfo {title} {Density-{Functional} {Theory} of {Atoms} and {Molecules}}}}\ (\bibinfo  {publisher} {Oxford University Press},\ \bibinfo {address} {New York},\ \bibinfo {year} {1989})\BibitemShut {NoStop}%
\bibitem [{\citenamefont {Dreizler}\ and\ \citenamefont {Gross}(1990)}]{Dreizler90}%
  \BibitemOpen
  \bibfield  {author} {\bibinfo {author} {\bibfnamefont {R.~M.}\ \bibnamefont {Dreizler}}\ and\ \bibinfo {author} {\bibfnamefont {E.~K.~U.}\ \bibnamefont {Gross}},\ }\href@noop {} {\emph {\bibinfo {title} {Density functional theory: an approach to the quantum many-body problem}}}\ (\bibinfo  {publisher} {Springer},\ \bibinfo {address} {Berlin Heidelberg},\ \bibinfo {year} {1990})\BibitemShut {NoStop}%
\bibitem [{\citenamefont {M.~Teale}\ \emph {et~al.}(2022)\citenamefont {M.~Teale}, \citenamefont {Helgaker}, \citenamefont {Savin}, \citenamefont {Adamo}, \citenamefont {Aradi}, \citenamefont {V.~Arbuznikov}, \citenamefont {W.~Ayers}, \citenamefont {Jan~Baerends}, \citenamefont {Barone}, \citenamefont {Calaminici}, \citenamefont {Cancès}, \citenamefont {A.~Carter}, \citenamefont {Kumar~Chattaraj}, \citenamefont {Chermette}, \citenamefont {Ciofini}, \citenamefont {Daniel~Crawford}, \citenamefont {Proft}, \citenamefont {F.~Dobson}, \citenamefont {Draxl}, \citenamefont {Frauenheim}, \citenamefont {Fromager}, \citenamefont {Fuentealba}, \citenamefont {Gagliardi}, \citenamefont {Galli}, \citenamefont {Gao}, \citenamefont {Geerlings}, \citenamefont {Gidopoulos}, \citenamefont {W.~Gill}, \citenamefont {Gori-Giorgi}, \citenamefont {Görling}, \citenamefont {Gould}, \citenamefont {Grimme}, \citenamefont {Gritsenko}, \citenamefont {Aagaard~Jensen}, \citenamefont {R.~Johnson}, \citenamefont {O.~Jones}, \citenamefont
  {Kaupp}, \citenamefont {M.~Köster}, \citenamefont {Kronik}, \citenamefont {I.~Krylov}, \citenamefont {Kvaal}, \citenamefont {Laestadius}, \citenamefont {Levy}, \citenamefont {Lewin}, \citenamefont {Liu}, \citenamefont {Loos}, \citenamefont {T.~Maitra}, \citenamefont {Neese}, \citenamefont {P.~Perdew}, \citenamefont {Pernal}, \citenamefont {Pernot}, \citenamefont {Piecuch}, \citenamefont {Rebolini}, \citenamefont {Reining}, \citenamefont {Romaniello}, \citenamefont {Ruzsinszky}, \citenamefont {R.~Salahub}, \citenamefont {Scheffler}, \citenamefont {Schwerdtfeger}, \citenamefont {N.~Staroverov}, \citenamefont {Sun}, \citenamefont {Tellgren}, \citenamefont {J.~Tozer}, \citenamefont {B.~Trickey}, \citenamefont {A.~Ullrich}, \citenamefont {Vela}, \citenamefont {Vignale}, \citenamefont {A.~Wesolowski}, \citenamefont {Xu},\ and\ \citenamefont {Yang}}]{Teale2228700}%
  \BibitemOpen
  \bibfield  {author} {\bibinfo {author} {\bibfnamefont {A.}~\bibnamefont {M.~Teale}}, \bibinfo {author} {\bibfnamefont {T.}~\bibnamefont {Helgaker}}, \bibinfo {author} {\bibfnamefont {A.}~\bibnamefont {Savin}}, \bibinfo {author} {\bibfnamefont {C.}~\bibnamefont {Adamo}}, \bibinfo {author} {\bibfnamefont {B.}~\bibnamefont {Aradi}}, \bibinfo {author} {\bibfnamefont {A.}~\bibnamefont {V.~Arbuznikov}}, \bibinfo {author} {\bibfnamefont {P.}~\bibnamefont {W.~Ayers}}, \bibinfo {author} {\bibfnamefont {E.}~\bibnamefont {Jan~Baerends}}, \bibinfo {author} {\bibfnamefont {V.}~\bibnamefont {Barone}}, \bibinfo {author} {\bibfnamefont {P.}~\bibnamefont {Calaminici}}, \bibinfo {author} {\bibfnamefont {E.}~\bibnamefont {Cancès}}, \bibinfo {author} {\bibfnamefont {E.}~\bibnamefont {A.~Carter}}, \bibinfo {author} {\bibfnamefont {P.}~\bibnamefont {Kumar~Chattaraj}}, \bibinfo {author} {\bibfnamefont {H.}~\bibnamefont {Chermette}}, \bibinfo {author} {\bibfnamefont {I.}~\bibnamefont {Ciofini}}, \bibinfo {author} {\bibfnamefont
  {T.}~\bibnamefont {Daniel~Crawford}}, \bibinfo {author} {\bibfnamefont {F.~D.}\ \bibnamefont {Proft}}, \bibinfo {author} {\bibfnamefont {J.}~\bibnamefont {F.~Dobson}}, \bibinfo {author} {\bibfnamefont {C.}~\bibnamefont {Draxl}}, \bibinfo {author} {\bibfnamefont {T.}~\bibnamefont {Frauenheim}}, \bibinfo {author} {\bibfnamefont {E.}~\bibnamefont {Fromager}}, \bibinfo {author} {\bibfnamefont {P.}~\bibnamefont {Fuentealba}}, \bibinfo {author} {\bibfnamefont {L.}~\bibnamefont {Gagliardi}}, \bibinfo {author} {\bibfnamefont {G.}~\bibnamefont {Galli}}, \bibinfo {author} {\bibfnamefont {J.}~\bibnamefont {Gao}}, \bibinfo {author} {\bibfnamefont {P.}~\bibnamefont {Geerlings}}, \bibinfo {author} {\bibfnamefont {N.}~\bibnamefont {Gidopoulos}}, \bibinfo {author} {\bibfnamefont {P.~M.}\ \bibnamefont {W.~Gill}}, \bibinfo {author} {\bibfnamefont {P.}~\bibnamefont {Gori-Giorgi}}, \bibinfo {author} {\bibfnamefont {A.}~\bibnamefont {Görling}}, \bibinfo {author} {\bibfnamefont {T.}~\bibnamefont {Gould}}, \bibinfo {author}
  {\bibfnamefont {S.}~\bibnamefont {Grimme}}, \bibinfo {author} {\bibfnamefont {O.}~\bibnamefont {Gritsenko}}, \bibinfo {author} {\bibfnamefont {H.~J.}\ \bibnamefont {Aagaard~Jensen}}, \bibinfo {author} {\bibfnamefont {E.}~\bibnamefont {R.~Johnson}}, \bibinfo {author} {\bibfnamefont {R.}~\bibnamefont {O.~Jones}}, \bibinfo {author} {\bibfnamefont {M.}~\bibnamefont {Kaupp}}, \bibinfo {author} {\bibfnamefont {A.}~\bibnamefont {M.~Köster}}, \bibinfo {author} {\bibfnamefont {L.}~\bibnamefont {Kronik}}, \bibinfo {author} {\bibfnamefont {A.}~\bibnamefont {I.~Krylov}}, \bibinfo {author} {\bibfnamefont {S.}~\bibnamefont {Kvaal}}, \bibinfo {author} {\bibfnamefont {A.}~\bibnamefont {Laestadius}}, \bibinfo {author} {\bibfnamefont {M.}~\bibnamefont {Levy}}, \bibinfo {author} {\bibfnamefont {M.}~\bibnamefont {Lewin}}, \bibinfo {author} {\bibfnamefont {S.}~\bibnamefont {Liu}}, \bibinfo {author} {\bibfnamefont {P.-F.}\ \bibnamefont {Loos}}, \bibinfo {author} {\bibfnamefont {N.}~\bibnamefont {T.~Maitra}}, \bibinfo {author}
  {\bibfnamefont {F.}~\bibnamefont {Neese}}, \bibinfo {author} {\bibfnamefont {J.}~\bibnamefont {P.~Perdew}}, \bibinfo {author} {\bibfnamefont {K.}~\bibnamefont {Pernal}}, \bibinfo {author} {\bibfnamefont {P.}~\bibnamefont {Pernot}}, \bibinfo {author} {\bibfnamefont {P.}~\bibnamefont {Piecuch}}, \bibinfo {author} {\bibfnamefont {E.}~\bibnamefont {Rebolini}}, \bibinfo {author} {\bibfnamefont {L.}~\bibnamefont {Reining}}, \bibinfo {author} {\bibfnamefont {P.}~\bibnamefont {Romaniello}}, \bibinfo {author} {\bibfnamefont {A.}~\bibnamefont {Ruzsinszky}}, \bibinfo {author} {\bibfnamefont {D.}~\bibnamefont {R.~Salahub}}, \bibinfo {author} {\bibfnamefont {M.}~\bibnamefont {Scheffler}}, \bibinfo {author} {\bibfnamefont {P.}~\bibnamefont {Schwerdtfeger}}, \bibinfo {author} {\bibfnamefont {V.}~\bibnamefont {N.~Staroverov}}, \bibinfo {author} {\bibfnamefont {J.}~\bibnamefont {Sun}}, \bibinfo {author} {\bibfnamefont {E.}~\bibnamefont {Tellgren}}, \bibinfo {author} {\bibfnamefont {D.}~\bibnamefont {J.~Tozer}}, \bibinfo
  {author} {\bibfnamefont {S.}~\bibnamefont {B.~Trickey}}, \bibinfo {author} {\bibfnamefont {C.}~\bibnamefont {A.~Ullrich}}, \bibinfo {author} {\bibfnamefont {A.}~\bibnamefont {Vela}}, \bibinfo {author} {\bibfnamefont {G.}~\bibnamefont {Vignale}}, \bibinfo {author} {\bibfnamefont {T.}~\bibnamefont {A.~Wesolowski}}, \bibinfo {author} {\bibfnamefont {X.}~\bibnamefont {Xu}},\ and\ \bibinfo {author} {\bibfnamefont {W.}~\bibnamefont {Yang}},\ }\bibfield  {title} {\bibinfo {title} {{{DFT}} exchange: Sharing perspectives on the workhorse of quantum chemistry and materials science},\ }\href {https://doi.org/10.1039/D2CP02827A} {\bibfield  {journal} {\bibinfo  {journal} {Physical Chemistry Chemical Physics}\ ,\ \bibinfo {pages} {28700}} (\bibinfo {year} {2022})}\BibitemShut {NoStop}%
\bibitem [{\citenamefont {Perdew}\ \emph {et~al.}(1996)\citenamefont {Perdew}, \citenamefont {Burke},\ and\ \citenamefont {Ernzerhof}}]{Perdew963865a}%
  \BibitemOpen
  \bibfield  {author} {\bibinfo {author} {\bibfnamefont {J.~P.}\ \bibnamefont {Perdew}}, \bibinfo {author} {\bibfnamefont {K.}~\bibnamefont {Burke}},\ and\ \bibinfo {author} {\bibfnamefont {M.}~\bibnamefont {Ernzerhof}},\ }\bibfield  {title} {\bibinfo {title} {Generalized gradient approximation made simple},\ }\href {https://doi.org/10.1103/PhysRevLett.77.3865} {\bibfield  {journal} {\bibinfo  {journal} {Physical Review Letters}\ }\textbf {\bibinfo {volume} {77}},\ \bibinfo {pages} {3865} (\bibinfo {year} {1996})}\BibitemShut {NoStop}%
\bibitem [{\citenamefont {Burke}(2012)}]{Burke12150901}%
  \BibitemOpen
  \bibfield  {author} {\bibinfo {author} {\bibfnamefont {K.}~\bibnamefont {Burke}},\ }\bibfield  {title} {\bibinfo {title} {Perspective on density functional theory},\ }\href {https://doi.org/10.1063/1.4704546} {\bibfield  {journal} {\bibinfo  {journal} {Journal of Chemical Physics}\ }\textbf {\bibinfo {volume} {136}},\ \bibinfo {pages} {150901} (\bibinfo {year} {2012})}\BibitemShut {NoStop}%
\bibitem [{\citenamefont {Cohen}\ \emph {et~al.}(2012)\citenamefont {Cohen}, \citenamefont {Mori-S{\'a}nchez},\ and\ \citenamefont {Yang}}]{Cohen12289}%
  \BibitemOpen
  \bibfield  {author} {\bibinfo {author} {\bibfnamefont {A.~J.}\ \bibnamefont {Cohen}}, \bibinfo {author} {\bibfnamefont {P.}~\bibnamefont {Mori-S{\'a}nchez}},\ and\ \bibinfo {author} {\bibfnamefont {W.}~\bibnamefont {Yang}},\ }\bibfield  {title} {\bibinfo {title} {Challenges for {Density} {Functional} {Theory}},\ }\href {https://doi.org/10.1021/cr200107z} {\bibfield  {journal} {\bibinfo  {journal} {Chemical Reviews}\ }\textbf {\bibinfo {volume} {112}},\ \bibinfo {pages} {289} (\bibinfo {year} {2012})}\BibitemShut {NoStop}%
\bibitem [{\citenamefont {Becke}(2014)}]{Becke1418A301}%
  \BibitemOpen
  \bibfield  {author} {\bibinfo {author} {\bibfnamefont {A.~D.}\ \bibnamefont {Becke}},\ }\bibfield  {title} {\bibinfo {title} {Perspective: {Fifty} years of density-functional theory in chemical physics},\ }\href {https://doi.org/10.1063/1.4869598} {\bibfield  {journal} {\bibinfo  {journal} {The Journal of Chemical Physics}\ }\textbf {\bibinfo {volume} {140}},\ \bibinfo {pages} {18A301} (\bibinfo {year} {2014})}\BibitemShut {NoStop}%
\bibitem [{\citenamefont {Sun}\ \emph {et~al.}(2015)\citenamefont {Sun}, \citenamefont {Ruzsinszky},\ and\ \citenamefont {Perdew}}]{Sun15036402a}%
  \BibitemOpen
  \bibfield  {author} {\bibinfo {author} {\bibfnamefont {J.}~\bibnamefont {Sun}}, \bibinfo {author} {\bibfnamefont {A.}~\bibnamefont {Ruzsinszky}},\ and\ \bibinfo {author} {\bibfnamefont {J.~P.}\ \bibnamefont {Perdew}},\ }\bibfield  {title} {\bibinfo {title} {Strongly {Constrained} and {Appropriately} {Normed} {Semilocal} {Density} {Functional}},\ }\href {https://doi.org/10.1103/PhysRevLett.115.036402} {\bibfield  {journal} {\bibinfo  {journal} {Physical Review Letters}\ }\textbf {\bibinfo {volume} {115}},\ \bibinfo {pages} {036402} (\bibinfo {year} {2015})}\BibitemShut {NoStop}%
\bibitem [{\citenamefont {Wang}\ \emph {et~al.}(2020)\citenamefont {Wang}, \citenamefont {Verma}, \citenamefont {Zhang}, \citenamefont {Li}, \citenamefont {Liu}, \citenamefont {Truhlar},\ and\ \citenamefont {He}}]{Wang202294}%
  \BibitemOpen
  \bibfield  {author} {\bibinfo {author} {\bibfnamefont {Y.}~\bibnamefont {Wang}}, \bibinfo {author} {\bibfnamefont {P.}~\bibnamefont {Verma}}, \bibinfo {author} {\bibfnamefont {L.}~\bibnamefont {Zhang}}, \bibinfo {author} {\bibfnamefont {Y.}~\bibnamefont {Li}}, \bibinfo {author} {\bibfnamefont {Z.}~\bibnamefont {Liu}}, \bibinfo {author} {\bibfnamefont {D.~G.}\ \bibnamefont {Truhlar}},\ and\ \bibinfo {author} {\bibfnamefont {X.}~\bibnamefont {He}},\ }\bibfield  {title} {\bibinfo {title} {M06-{SX} screened-exchange density functional for chemistry and solid-state physics},\ }\href {https://doi.org/10/gmt6vm} {\bibfield  {journal} {\bibinfo  {journal} {Proceedings of the National Academy of Sciences}\ }\textbf {\bibinfo {volume} {117}},\ \bibinfo {pages} {2294} (\bibinfo {year} {2020})}\BibitemShut {NoStop}%
\bibitem [{\citenamefont {Mardirossian}\ and\ \citenamefont {Head-Gordon}(2016)}]{Mardirossian16214110}%
  \BibitemOpen
  \bibfield  {author} {\bibinfo {author} {\bibfnamefont {N.}~\bibnamefont {Mardirossian}}\ and\ \bibinfo {author} {\bibfnamefont {M.}~\bibnamefont {Head-Gordon}},\ }\bibfield  {title} {\bibinfo {title} {omega {B97M}-{V}: {A} combinatorially optimized, range-separated hybrid, meta-{GGA} density functional with {VV10} nonlocal correlation},\ }\href {https://doi.org/10.1063/1.4952647} {\bibfield  {journal} {\bibinfo  {journal} {Journal of Chemical Physics}\ }\textbf {\bibinfo {volume} {144}},\ \bibinfo {pages} {214110} (\bibinfo {year} {2016})}\BibitemShut {NoStop}%
\bibitem [{\citenamefont {Perdew}\ \emph {et~al.}(1982)\citenamefont {Perdew}, \citenamefont {Parr}, \citenamefont {Levy},\ and\ \citenamefont {Balduz}}]{Perdew821691a}%
  \BibitemOpen
  \bibfield  {author} {\bibinfo {author} {\bibfnamefont {J.}~\bibnamefont {Perdew}}, \bibinfo {author} {\bibfnamefont {R.}~\bibnamefont {Parr}}, \bibinfo {author} {\bibfnamefont {M.}~\bibnamefont {Levy}},\ and\ \bibinfo {author} {\bibfnamefont {J.}~\bibnamefont {Balduz}},\ }\bibfield  {title} {\bibinfo {title} {Density-{Functional} {Theory} for {Fractional} {Particle} {Number} - {Derivative} {Discontinuities} of the {Energy}},\ }\href {https://doi.org/10.1103/PhysRevLett.49.1691} {\bibfield  {journal} {\bibinfo  {journal} {Physical Review Letters}\ }\textbf {\bibinfo {volume} {49}},\ \bibinfo {pages} {1691} (\bibinfo {year} {1982})}\BibitemShut {NoStop}%
\bibitem [{\citenamefont {Zhang}\ and\ \citenamefont {Yang}(2000)}]{Zhang00346}%
  \BibitemOpen
  \bibfield  {author} {\bibinfo {author} {\bibfnamefont {Y.}~\bibnamefont {Zhang}}\ and\ \bibinfo {author} {\bibfnamefont {W.}~\bibnamefont {Yang}},\ }\bibfield  {title} {\bibinfo {title} {Perspective on "{Density}-functional theory for fractional particle number: derivative discontinuities of the energy"},\ }\href {https://doi.org/10.1007/s002149900021} {\bibfield  {journal} {\bibinfo  {journal} {Theoretical Chemistry Accounts: Theory, Computation, and Modeling (Theoretica Chimica Acta)}\ }\textbf {\bibinfo {volume} {103}},\ \bibinfo {pages} {346} (\bibinfo {year} {2000})}\BibitemShut {NoStop}%
\bibitem [{\citenamefont {Yang}\ \emph {et~al.}(2000)\citenamefont {Yang}, \citenamefont {Zhang},\ and\ \citenamefont {Ayers}}]{Yang005172}%
  \BibitemOpen
  \bibfield  {author} {\bibinfo {author} {\bibfnamefont {W.}~\bibnamefont {Yang}}, \bibinfo {author} {\bibfnamefont {Y.}~\bibnamefont {Zhang}},\ and\ \bibinfo {author} {\bibfnamefont {P.~W.}\ \bibnamefont {Ayers}},\ }\bibfield  {title} {\bibinfo {title} {Degenerate ground states and a fractional number of electrons in density and reduced density matrix functional theory},\ }\href {https://journals.aps.org/prl/abstract/10.1103/PhysRevLett.84.5172} {\bibfield  {journal} {\bibinfo  {journal} {Physical Review Letters}\ }\textbf {\bibinfo {volume} {84}},\ \bibinfo {pages} {5172} (\bibinfo {year} {2000})}\BibitemShut {NoStop}%
\bibitem [{\citenamefont {Cohen}\ \emph {et~al.}(2008{\natexlab{a}})\citenamefont {Cohen}, \citenamefont {Mori-S{\'a}nchez},\ and\ \citenamefont {Yang}}]{Cohen08121104}%
  \BibitemOpen
  \bibfield  {author} {\bibinfo {author} {\bibfnamefont {A.~J.}\ \bibnamefont {Cohen}}, \bibinfo {author} {\bibfnamefont {P.}~\bibnamefont {Mori-S{\'a}nchez}},\ and\ \bibinfo {author} {\bibfnamefont {W.}~\bibnamefont {Yang}},\ }\bibfield  {title} {\bibinfo {title} {Fractional spins and static correlation error in density functional theory},\ }\href {https://doi.org/10.1063/1.2987202} {\bibfield  {journal} {\bibinfo  {journal} {The Journal of Chemical Physics}\ }\textbf {\bibinfo {volume} {129}},\ \bibinfo {pages} {121104} (\bibinfo {year} {2008}{\natexlab{a}})}\BibitemShut {NoStop}%
\bibitem [{\citenamefont {Mori-S{\'a}nchez}\ \emph {et~al.}(2009)\citenamefont {Mori-S{\'a}nchez}, \citenamefont {Cohen},\ and\ \citenamefont {Yang}}]{Mori-Sanchez09066403a}%
  \BibitemOpen
  \bibfield  {author} {\bibinfo {author} {\bibfnamefont {P.}~\bibnamefont {Mori-S{\'a}nchez}}, \bibinfo {author} {\bibfnamefont {A.~J.}\ \bibnamefont {Cohen}},\ and\ \bibinfo {author} {\bibfnamefont {W.}~\bibnamefont {Yang}},\ }\bibfield  {title} {\bibinfo {title} {Discontinuous {Nature} of the {Exchange}-{Correlation} {Functional} in {Strongly} {Correlated} {Systems}},\ }\href {https://doi.org/10.1103/PhysRevLett.102.066403} {\bibfield  {journal} {\bibinfo  {journal} {Physical Review Letters}\ }\textbf {\bibinfo {volume} {102}},\ \bibinfo {pages} {066403} (\bibinfo {year} {2009})}\BibitemShut {NoStop}%
\bibitem [{\citenamefont {Mori-S{\'a}nchez}\ \emph {et~al.}(2008)\citenamefont {Mori-S{\'a}nchez}, \citenamefont {Cohen},\ and\ \citenamefont {Yang}}]{Mori-Sanchez08146401}%
  \BibitemOpen
  \bibfield  {author} {\bibinfo {author} {\bibfnamefont {P.}~\bibnamefont {Mori-S{\'a}nchez}}, \bibinfo {author} {\bibfnamefont {A.~J.}\ \bibnamefont {Cohen}},\ and\ \bibinfo {author} {\bibfnamefont {W.}~\bibnamefont {Yang}},\ }\bibfield  {title} {\bibinfo {title} {Localization and {Delocalization} {Errors} in {Density} {Functional} {Theory} and {Implications} for {Band}-{Gap} {Prediction}},\ }\href {https://doi.org/10/fw5jz9} {\bibfield  {journal} {\bibinfo  {journal} {Physical Review Letters}\ }\textbf {\bibinfo {volume} {100}},\ \bibinfo {pages} {146401} (\bibinfo {year} {2008})}\BibitemShut {NoStop}%
\bibitem [{\citenamefont {Cohen}\ \emph {et~al.}(2008{\natexlab{b}})\citenamefont {Cohen}, \citenamefont {Mori-Sanchez},\ and\ \citenamefont {Yang}}]{Cohen08792}%
  \BibitemOpen
  \bibfield  {author} {\bibinfo {author} {\bibfnamefont {A.~J.}\ \bibnamefont {Cohen}}, \bibinfo {author} {\bibfnamefont {P.}~\bibnamefont {Mori-Sanchez}},\ and\ \bibinfo {author} {\bibfnamefont {W.}~\bibnamefont {Yang}},\ }\bibfield  {title} {\bibinfo {title} {Insights into {Current} {Limitations} of {Density} {Functional} {Theory}},\ }\href {https://doi.org/10.1126/science.1158722} {\bibfield  {journal} {\bibinfo  {journal} {Science}\ }\textbf {\bibinfo {volume} {321}},\ \bibinfo {pages} {792} (\bibinfo {year} {2008}{\natexlab{b}})}\BibitemShut {NoStop}%
\bibitem [{\citenamefont {Bryenton}\ \emph {et~al.}(2022)\citenamefont {Bryenton}, \citenamefont {Adeleke}, \citenamefont {Dale},\ and\ \citenamefont {Johnson}}]{Bryenton22e1631a}%
  \BibitemOpen
  \bibfield  {author} {\bibinfo {author} {\bibfnamefont {K.~R.}\ \bibnamefont {Bryenton}}, \bibinfo {author} {\bibfnamefont {A.~A.}\ \bibnamefont {Adeleke}}, \bibinfo {author} {\bibfnamefont {S.~G.}\ \bibnamefont {Dale}},\ and\ \bibinfo {author} {\bibfnamefont {E.~R.}\ \bibnamefont {Johnson}},\ }\bibfield  {title} {\bibinfo {title} {Delocalization error: {The} greatest outstanding challenge in density-functional theory},\ }\href {https://doi.org/10.1002/wcms.1631} {\bibfield  {journal} {\bibinfo  {journal} {Wiley Interdisciplinary Reviews-Computational Molecular Science}\ }\textbf {\bibinfo {volume} {13}},\ \bibinfo {pages} {e1631} (\bibinfo {year} {2022})}\BibitemShut {NoStop}%
\bibitem [{\citenamefont {Mei}\ \emph {et~al.}(2022)\citenamefont {Mei}, \citenamefont {Yu}, \citenamefont {Chen}, \citenamefont {Su},\ and\ \citenamefont {Yang}}]{Mei22acs.jctc.1c01058}%
  \BibitemOpen
  \bibfield  {author} {\bibinfo {author} {\bibfnamefont {Y.}~\bibnamefont {Mei}}, \bibinfo {author} {\bibfnamefont {J.}~\bibnamefont {Yu}}, \bibinfo {author} {\bibfnamefont {Z.}~\bibnamefont {Chen}}, \bibinfo {author} {\bibfnamefont {N.~Q.}\ \bibnamefont {Su}},\ and\ \bibinfo {author} {\bibfnamefont {W.}~\bibnamefont {Yang}},\ }\bibfield  {title} {\bibinfo {title} {{LibSC}: {Library} for {Scaling} {Correction} {Methods} in {Density} {Functional} {Theory}},\ }\href {https://doi.org/10/gpb64w} {\bibfield  {journal} {\bibinfo  {journal} {Journal of Chemical Theory and Computation}\ }\textbf {\bibinfo {volume} {18}},\ \bibinfo {pages} {840} (\bibinfo {year} {2022})}\BibitemShut {NoStop}%
\bibitem [{\citenamefont {Perdew}\ and\ \citenamefont {Zunger}(1981)}]{Perdew815048}%
  \BibitemOpen
  \bibfield  {author} {\bibinfo {author} {\bibfnamefont {J.}~\bibnamefont {Perdew}}\ and\ \bibinfo {author} {\bibfnamefont {A.}~\bibnamefont {Zunger}},\ }\bibfield  {title} {\bibinfo {title} {Self-{Interaction} {Correction} to {Density}-{Functional} {Approximations} for {Many}-{Electron} {Systems}},\ }\href {https://doi.org/10.1103/PhysRevB.23.5048} {\bibfield  {journal} {\bibinfo  {journal} {Physical Review B}\ }\textbf {\bibinfo {volume} {23}},\ \bibinfo {pages} {5048} (\bibinfo {year} {1981})}\BibitemShut {NoStop}%
\bibitem [{\citenamefont {Pederson}\ \emph {et~al.}(2014)\citenamefont {Pederson}, \citenamefont {Ruzsinszky},\ and\ \citenamefont {Perdew}}]{Pederson14121103a}%
  \BibitemOpen
  \bibfield  {author} {\bibinfo {author} {\bibfnamefont {M.~R.}\ \bibnamefont {Pederson}}, \bibinfo {author} {\bibfnamefont {A.}~\bibnamefont {Ruzsinszky}},\ and\ \bibinfo {author} {\bibfnamefont {J.~P.}\ \bibnamefont {Perdew}},\ }\bibfield  {title} {\bibinfo {title} {Communication: {Self}-interaction correction with unitary invariance in density functional theory},\ }\href {https://doi.org/10.1063/1.4869581} {\bibfield  {journal} {\bibinfo  {journal} {Journal of Chemical Physics}\ }\textbf {\bibinfo {volume} {140}},\ \bibinfo {pages} {121103} (\bibinfo {year} {2014})}\BibitemShut {NoStop}%
\bibitem [{\citenamefont {Savin}(1996)}]{Savin96327}%
  \BibitemOpen
  \bibfield  {author} {\bibinfo {author} {\bibfnamefont {A.}~\bibnamefont {Savin}},\ }\bibfield  {title} {\bibinfo {title} {On degeneracy, near-degeneracy and density functional theory},\ }in\ \href {https://doi.org/10.1016/S1380-7323(96)80091-4} {\emph {\bibinfo {booktitle} {Theoretical and {{Computational Chemistry}}}}},\ Vol.~\bibinfo {volume} {4}\ (\bibinfo  {publisher} {Elsevier},\ \bibinfo {year} {1996})\ pp.\ \bibinfo {pages} {327--357}\BibitemShut {NoStop}%
\bibitem [{\citenamefont {Iikura}\ \emph {et~al.}(2001)\citenamefont {Iikura}, \citenamefont {Tsuneda}, \citenamefont {Yanai},\ and\ \citenamefont {Hirao}}]{Iikura013540}%
  \BibitemOpen
  \bibfield  {author} {\bibinfo {author} {\bibfnamefont {H.}~\bibnamefont {Iikura}}, \bibinfo {author} {\bibfnamefont {T.}~\bibnamefont {Tsuneda}}, \bibinfo {author} {\bibfnamefont {T.}~\bibnamefont {Yanai}},\ and\ \bibinfo {author} {\bibfnamefont {K.}~\bibnamefont {Hirao}},\ }\bibfield  {title} {\bibinfo {title} {A long-range correction scheme for generalized-gradient-approximation exchange functionals},\ }\href {https://doi.org/10/b5v8q9} {\bibfield  {journal} {\bibinfo  {journal} {The Journal of Chemical Physics}\ }\textbf {\bibinfo {volume} {115}},\ \bibinfo {pages} {3540} (\bibinfo {year} {2001})}\BibitemShut {NoStop}%
\bibitem [{\citenamefont {Song}\ \emph {et~al.}(2009)\citenamefont {Song}, \citenamefont {Watson},\ and\ \citenamefont {Hirao}}]{Song09144108}%
  \BibitemOpen
  \bibfield  {author} {\bibinfo {author} {\bibfnamefont {J.-W.}\ \bibnamefont {Song}}, \bibinfo {author} {\bibfnamefont {M.~A.}\ \bibnamefont {Watson}},\ and\ \bibinfo {author} {\bibfnamefont {K.}~\bibnamefont {Hirao}},\ }\bibfield  {title} {\bibinfo {title} {An improved long-range corrected hybrid functional with vanishing {Hartree}-{Fock} exchange at zero interelectronic distance, {LC2gau}-{BOP}},\ }\href {https://doi.org/10.1063/1.3243819} {\bibfield  {journal} {\bibinfo  {journal} {Journal of Chemical Physics}\ }\textbf {\bibinfo {volume} {131}},\ \bibinfo {pages} {144108} (\bibinfo {year} {2009})}\BibitemShut {NoStop}%
\bibitem [{\citenamefont {Wagle}\ \emph {et~al.}(2021)\citenamefont {Wagle}, \citenamefont {Santra}, \citenamefont {Bhattarai}, \citenamefont {Shahi}, \citenamefont {Pederson}, \citenamefont {Jackson},\ and\ \citenamefont {Perdew}}]{Wagle21094302c}%
  \BibitemOpen
  \bibfield  {author} {\bibinfo {author} {\bibfnamefont {K.}~\bibnamefont {Wagle}}, \bibinfo {author} {\bibfnamefont {B.}~\bibnamefont {Santra}}, \bibinfo {author} {\bibfnamefont {P.}~\bibnamefont {Bhattarai}}, \bibinfo {author} {\bibfnamefont {C.}~\bibnamefont {Shahi}}, \bibinfo {author} {\bibfnamefont {M.~R.}\ \bibnamefont {Pederson}}, \bibinfo {author} {\bibfnamefont {K.~A.}\ \bibnamefont {Jackson}},\ and\ \bibinfo {author} {\bibfnamefont {J.~P.}\ \bibnamefont {Perdew}},\ }\bibfield  {title} {\bibinfo {title} {Self-interaction correction in water{\textendash}ion clusters},\ }\href {https://doi.org/10/gh62js} {\bibfield  {journal} {\bibinfo  {journal} {The Journal of Chemical Physics}\ }\textbf {\bibinfo {volume} {154}},\ \bibinfo {pages} {094302} (\bibinfo {year} {2021})}\BibitemShut {NoStop}%
\bibitem [{\citenamefont {Baer}\ \emph {et~al.}(2010)\citenamefont {Baer}, \citenamefont {Livshits},\ and\ \citenamefont {Salzner}}]{Baer1085}%
  \BibitemOpen
  \bibfield  {author} {\bibinfo {author} {\bibfnamefont {R.}~\bibnamefont {Baer}}, \bibinfo {author} {\bibfnamefont {E.}~\bibnamefont {Livshits}},\ and\ \bibinfo {author} {\bibfnamefont {U.}~\bibnamefont {Salzner}},\ }\bibfield  {title} {\bibinfo {title} {Tuned range-separated hybrids in density functional theory},\ }\href {https://doi.org/https://doi.org/10.1146/annurev.physchem.012809.103321} {\bibfield  {journal} {\bibinfo  {journal} {Annual Review of Physical Chemistry}\ }\textbf {\bibinfo {volume} {61}},\ \bibinfo {pages} {85} (\bibinfo {year} {2010})}\BibitemShut {NoStop}%
\bibitem [{\citenamefont {Wing}\ \emph {et~al.}(2021)\citenamefont {Wing}, \citenamefont {Ohad}, \citenamefont {Haber}, \citenamefont {Filip}, \citenamefont {Gant}, \citenamefont {Neaton},\ and\ \citenamefont {Kronik}}]{Wing21}%
  \BibitemOpen
  \bibfield  {author} {\bibinfo {author} {\bibfnamefont {D.}~\bibnamefont {Wing}}, \bibinfo {author} {\bibfnamefont {G.}~\bibnamefont {Ohad}}, \bibinfo {author} {\bibfnamefont {J.~B.}\ \bibnamefont {Haber}}, \bibinfo {author} {\bibfnamefont {M.~R.}\ \bibnamefont {Filip}}, \bibinfo {author} {\bibfnamefont {S.~E.}\ \bibnamefont {Gant}}, \bibinfo {author} {\bibfnamefont {J.~B.}\ \bibnamefont {Neaton}},\ and\ \bibinfo {author} {\bibfnamefont {L.}~\bibnamefont {Kronik}},\ }\bibfield  {title} {\bibinfo {title} {Band gaps of crystalline solids from {Wannier}-localization{\textendash}based optimal tuning of a screened range-separated hybrid functional},\ }\bibfield  {journal} {\bibinfo  {journal} {Proceedings of the National Academy of Sciences}\ }\textbf {\bibinfo {volume} {118}},\ \href {https://doi.org/10/gmt8wp} {10/gmt8wp} (\bibinfo {year} {2021})\BibitemShut {NoStop}%
\bibitem [{\citenamefont {Su}\ \emph {et~al.}(2014)\citenamefont {Su}, \citenamefont {Yang}, \citenamefont {{Mori-S{\'a}nchez}},\ and\ \citenamefont {Xu}}]{Su149201b}%
  \BibitemOpen
  \bibfield  {author} {\bibinfo {author} {\bibfnamefont {N.~Q.}\ \bibnamefont {Su}}, \bibinfo {author} {\bibfnamefont {W.}~\bibnamefont {Yang}}, \bibinfo {author} {\bibfnamefont {P.}~\bibnamefont {{Mori-S{\'a}nchez}}},\ and\ \bibinfo {author} {\bibfnamefont {X.}~\bibnamefont {Xu}},\ }\bibfield  {title} {\bibinfo {title} {Fractional {{Charge Behavior}} and {{Band Gap Predictions}} with the {{XYG3 Type}} of {{Doubly Hybrid Density Functionals}}},\ }\href {https://doi.org/10/f6kp57} {\bibfield  {journal} {\bibinfo  {journal} {J. Phys. Chem. A}\ }\textbf {\bibinfo {volume} {118}},\ \bibinfo {pages} {9201} (\bibinfo {year} {2014})}\BibitemShut {NoStop}%
\bibitem [{\citenamefont {Su}\ and\ \citenamefont {Xu}(2016)}]{Su162285a}%
  \BibitemOpen
  \bibfield  {author} {\bibinfo {author} {\bibfnamefont {N.~Q.}\ \bibnamefont {Su}}\ and\ \bibinfo {author} {\bibfnamefont {X.}~\bibnamefont {Xu}},\ }\bibfield  {title} {\bibinfo {title} {Second-{{Order Perturbation Theory}} for {{Fractional Occupation Systems}}: {{Applications}} to {{Ionization Potential}} and {{Electron Affinity Calculations}}},\ }\href {https://doi.org/10/gfz4m8} {\bibfield  {journal} {\bibinfo  {journal} {J. Chem. Theory Comput.}\ }\textbf {\bibinfo {volume} {12}},\ \bibinfo {pages} {2285} (\bibinfo {year} {2016})}\BibitemShut {NoStop}%
\bibitem [{\citenamefont {Li}\ and\ \citenamefont {Yang}()}]{Li244876}%
  \BibitemOpen
  \bibfield  {author} {\bibinfo {author} {\bibfnamefont {J.}~\bibnamefont {Li}}\ and\ \bibinfo {author} {\bibfnamefont {W.}~\bibnamefont {Yang}},\ }\bibfield  {title} {\bibinfo {title} {Chemical {{Potentials}} and the {{One-Electron Hamiltonian}} of the {{Second-Order Perturbation Theory}} from the {{Functional Derivative Approach}}},\ }\href {https://doi.org/10.1021/acs.jpca.4c01574} {\bibfield  {journal} {\bibinfo  {journal} {J. Phys. Chem. A}\ }\textbf {\bibinfo {volume} {128}},\ \bibinfo {pages} {4876}}\BibitemShut {NoStop}%
\bibitem [{\citenamefont {Anisimov}\ and\ \citenamefont {Kozhevnikov}(2005)}]{Anisimov05075125}%
  \BibitemOpen
  \bibfield  {author} {\bibinfo {author} {\bibfnamefont {V.~I.}\ \bibnamefont {Anisimov}}\ and\ \bibinfo {author} {\bibfnamefont {A.~V.}\ \bibnamefont {Kozhevnikov}},\ }\bibfield  {title} {\bibinfo {title} {Transition state method and {Wannier} functions},\ }\href {https://doi.org/10.1103/PhysRevB.72.075125} {\bibfield  {journal} {\bibinfo  {journal} {Physical Review B}\ }\textbf {\bibinfo {volume} {72}},\ \bibinfo {pages} {075125} (\bibinfo {year} {2005})}\BibitemShut {NoStop}%
\bibitem [{\citenamefont {Ma}\ and\ \citenamefont {Wang}(2016)}]{Ma1624924}%
  \BibitemOpen
  \bibfield  {author} {\bibinfo {author} {\bibfnamefont {J.}~\bibnamefont {Ma}}\ and\ \bibinfo {author} {\bibfnamefont {L.-W.}\ \bibnamefont {Wang}},\ }\bibfield  {title} {\bibinfo {title} {Using {Wannier} functions to improve solid band gap predictions in density functional theory},\ }\href {https://doi.org/10.1038/srep24924} {\bibfield  {journal} {\bibinfo  {journal} {Scientific Reports}\ }\textbf {\bibinfo {volume} {6}},\ \bibinfo {pages} {24924} (\bibinfo {year} {2016})}\BibitemShut {NoStop}%
\bibitem [{\citenamefont {Dabo}\ \emph {et~al.}(2010)\citenamefont {Dabo}, \citenamefont {Ferretti}, \citenamefont {Poilvert}, \citenamefont {Li}, \citenamefont {Marzari},\ and\ \citenamefont {Cococcioni}}]{Dabo10115121}%
  \BibitemOpen
  \bibfield  {author} {\bibinfo {author} {\bibfnamefont {I.}~\bibnamefont {Dabo}}, \bibinfo {author} {\bibfnamefont {A.}~\bibnamefont {Ferretti}}, \bibinfo {author} {\bibfnamefont {N.}~\bibnamefont {Poilvert}}, \bibinfo {author} {\bibfnamefont {Y.}~\bibnamefont {Li}}, \bibinfo {author} {\bibfnamefont {N.}~\bibnamefont {Marzari}},\ and\ \bibinfo {author} {\bibfnamefont {M.}~\bibnamefont {Cococcioni}},\ }\bibfield  {title} {\bibinfo {title} {Koopmans' condition for density-functional theory},\ }\href {https://doi.org/10.1103/PhysRevB.82.115121} {\bibfield  {journal} {\bibinfo  {journal} {Physical Review B}\ }\textbf {\bibinfo {volume} {82}},\ \bibinfo {pages} {115121} (\bibinfo {year} {2010})}\BibitemShut {NoStop}%
\bibitem [{\citenamefont {Colonna}\ \emph {et~al.}(2022{\natexlab{a}})\citenamefont {Colonna}, \citenamefont {De~Gennaro}, \citenamefont {Linscott},\ and\ \citenamefont {Marzari}}]{Colonna22}%
  \BibitemOpen
  \bibfield  {author} {\bibinfo {author} {\bibfnamefont {N.}~\bibnamefont {Colonna}}, \bibinfo {author} {\bibfnamefont {R.}~\bibnamefont {De~Gennaro}}, \bibinfo {author} {\bibfnamefont {E.}~\bibnamefont {Linscott}},\ and\ \bibinfo {author} {\bibfnamefont {N.}~\bibnamefont {Marzari}},\ }\bibfield  {title} {\bibinfo {title} {Koopmans spectral functionals in periodic-boundary conditions},\ }\href@noop {} {\bibfield  {journal} {\bibinfo  {journal} {Journal of Chemical Theory and Computation}\ }\textbf {\bibinfo {volume} {18}} (\bibinfo {year} {2022}{\natexlab{a}})}\BibitemShut {NoStop}%
\bibitem [{\citenamefont {Cohen}\ \emph {et~al.}(2007)\citenamefont {Cohen}, \citenamefont {Mori-S{\'a}nchez},\ and\ \citenamefont {Yang}}]{Cohen07191109}%
  \BibitemOpen
  \bibfield  {author} {\bibinfo {author} {\bibfnamefont {A.~J.}\ \bibnamefont {Cohen}}, \bibinfo {author} {\bibfnamefont {P.}~\bibnamefont {Mori-S{\'a}nchez}},\ and\ \bibinfo {author} {\bibfnamefont {W.}~\bibnamefont {Yang}},\ }\bibfield  {title} {\bibinfo {title} {Development of exchange-correlation functionals with minimal many-electron self-interaction error},\ }\href {https://doi.org/10.1063/1.2741248} {\bibfield  {journal} {\bibinfo  {journal} {The Journal of Chemical Physics}\ }\textbf {\bibinfo {volume} {126}},\ \bibinfo {pages} {191109} (\bibinfo {year} {2007})}\BibitemShut {NoStop}%
\bibitem [{\citenamefont {Zheng}\ \emph {et~al.}(2011)\citenamefont {Zheng}, \citenamefont {Cohen}, \citenamefont {Mori-S{\'a}nchez}, \citenamefont {Hu},\ and\ \citenamefont {Yang}}]{Zheng11}%
  \BibitemOpen
  \bibfield  {author} {\bibinfo {author} {\bibfnamefont {X.}~\bibnamefont {Zheng}}, \bibinfo {author} {\bibfnamefont {A.~J.}\ \bibnamefont {Cohen}}, \bibinfo {author} {\bibfnamefont {P.}~\bibnamefont {Mori-S{\'a}nchez}}, \bibinfo {author} {\bibfnamefont {X.}~\bibnamefont {Hu}},\ and\ \bibinfo {author} {\bibfnamefont {W.}~\bibnamefont {Yang}},\ }\bibfield  {title} {\bibinfo {title} {Improving {Band} {Gap} {Prediction} in {Density} {Functional} {Theory} from {Molecules} to {Solids}},\ }\bibfield  {journal} {\bibinfo  {journal} {Physical Review Letters}\ }\textbf {\bibinfo {volume} {107}},\ \href {https://doi.org/10.1103/PhysRevLett.107.026403} {10.1103/PhysRevLett.107.026403} (\bibinfo {year} {2011})\BibitemShut {NoStop}%
\bibitem [{\citenamefont {Li}\ \emph {et~al.}(2015)\citenamefont {Li}, \citenamefont {Zheng}, \citenamefont {Cohen}, \citenamefont {Mori-S{\'a}nchez},\ and\ \citenamefont {Yang}}]{Li15053001}%
  \BibitemOpen
  \bibfield  {author} {\bibinfo {author} {\bibfnamefont {C.}~\bibnamefont {Li}}, \bibinfo {author} {\bibfnamefont {X.}~\bibnamefont {Zheng}}, \bibinfo {author} {\bibfnamefont {A.~J.}\ \bibnamefont {Cohen}}, \bibinfo {author} {\bibfnamefont {P.}~\bibnamefont {Mori-S{\'a}nchez}},\ and\ \bibinfo {author} {\bibfnamefont {W.}~\bibnamefont {Yang}},\ }\bibfield  {title} {\bibinfo {title} {Local {Scaling} {Correction} for {Reducing} {Delocalization} {Error} in {Density} {Functional} {Approximations}},\ }\href {https://doi.org/10.1103/PhysRevLett.114.053001} {\bibfield  {journal} {\bibinfo  {journal} {Physical Review Letters}\ }\textbf {\bibinfo {volume} {114}},\ \bibinfo {pages} {053001} (\bibinfo {year} {2015})}\BibitemShut {NoStop}%
\bibitem [{\citenamefont {Li}\ \emph {et~al.}(2018)\citenamefont {Li}, \citenamefont {Zheng}, \citenamefont {Su},\ and\ \citenamefont {Yang}}]{Li18203}%
  \BibitemOpen
  \bibfield  {author} {\bibinfo {author} {\bibfnamefont {C.}~\bibnamefont {Li}}, \bibinfo {author} {\bibfnamefont {X.}~\bibnamefont {Zheng}}, \bibinfo {author} {\bibfnamefont {N.~Q.}\ \bibnamefont {Su}},\ and\ \bibinfo {author} {\bibfnamefont {W.}~\bibnamefont {Yang}},\ }\bibfield  {title} {\bibinfo {title} {Localized orbital scaling correction for systematic elimination of delocalization error in density functional approximations},\ }\href {https://doi.org/10.1093/nsr/nwx111} {\bibfield  {journal} {\bibinfo  {journal} {National Science Review}\ }\textbf {\bibinfo {volume} {5}},\ \bibinfo {pages} {203} (\bibinfo {year} {2018})}\BibitemShut {NoStop}%
\bibitem [{\citenamefont {Su}\ \emph {et~al.}(2020{\natexlab{a}})\citenamefont {Su}, \citenamefont {Mahler},\ and\ \citenamefont {Yang}}]{Su201528}%
  \BibitemOpen
  \bibfield  {author} {\bibinfo {author} {\bibfnamefont {N.~Q.}\ \bibnamefont {Su}}, \bibinfo {author} {\bibfnamefont {A.}~\bibnamefont {Mahler}},\ and\ \bibinfo {author} {\bibfnamefont {W.}~\bibnamefont {Yang}},\ }\bibfield  {title} {\bibinfo {title} {Preserving {Symmetry} and {Degeneracy} in the {Localized} {Orbital} {Scaling} {Correction} {Approach}},\ }\href {https://doi.org/10/ggnjbs} {\bibfield  {journal} {\bibinfo  {journal} {The Journal of Physical Chemistry Letters}\ }\textbf {\bibinfo {volume} {11}},\ \bibinfo {pages} {1528} (\bibinfo {year} {2020}{\natexlab{a}})}\BibitemShut {NoStop}%
\bibitem [{\citenamefont {Janak}(1978)}]{Janak787165}%
  \BibitemOpen
  \bibfield  {author} {\bibinfo {author} {\bibfnamefont {J.~F.}\ \bibnamefont {Janak}},\ }\bibfield  {title} {\bibinfo {title} {Proof that $\cfrac{\partial e}{\partial n_i} = \varepsilon_i$ in density-functional theory},\ }\href {https://doi.org/10/dw6bqt} {\bibfield  {journal} {\bibinfo  {journal} {Physical Review B}\ }\textbf {\bibinfo {volume} {18}},\ \bibinfo {pages} {7165} (\bibinfo {year} {1978})}\BibitemShut {NoStop}%
\bibitem [{\citenamefont {Morrell}\ \emph {et~al.}(1975)\citenamefont {Morrell}, \citenamefont {Parr},\ and\ \citenamefont {Levy}}]{Morrell75549a}%
  \BibitemOpen
  \bibfield  {author} {\bibinfo {author} {\bibfnamefont {M.~M.}\ \bibnamefont {Morrell}}, \bibinfo {author} {\bibfnamefont {R.~G.}\ \bibnamefont {Parr}},\ and\ \bibinfo {author} {\bibfnamefont {M.}~\bibnamefont {Levy}},\ }\bibfield  {title} {\bibinfo {title} {Calculation of ionization potentials from density matrices and natural functions, and the long-range behavior of natural orbitals and electron density},\ }\href@noop {} {\bibfield  {journal} {\bibinfo  {journal} {The Journal of Chemical Physics}\ }\textbf {\bibinfo {volume} {62}},\ \bibinfo {pages} {549} (\bibinfo {year} {1975})}\BibitemShut {NoStop}%
\bibitem [{\citenamefont {Perdew}\ and\ \citenamefont {Levy}(1997)}]{Perdew9716021b}%
  \BibitemOpen
  \bibfield  {author} {\bibinfo {author} {\bibfnamefont {J.~P.}\ \bibnamefont {Perdew}}\ and\ \bibinfo {author} {\bibfnamefont {M.}~\bibnamefont {Levy}},\ }\bibfield  {title} {\bibinfo {title} {Comment on ``{Significance} of the highest occupied {Kohn}-{Sham} eigenvalue''},\ }\href {https://doi.org/10.1103/PhysRevB.56.16021} {\bibfield  {journal} {\bibinfo  {journal} {Physical Review B}\ }\textbf {\bibinfo {volume} {56}},\ \bibinfo {pages} {16021} (\bibinfo {year} {1997})}\BibitemShut {NoStop}%
\bibitem [{\citenamefont {Tozer}\ and\ \citenamefont {Handy}(2000)}]{Tozer002117b}%
  \BibitemOpen
  \bibfield  {author} {\bibinfo {author} {\bibfnamefont {D.~J.}\ \bibnamefont {Tozer}}\ and\ \bibinfo {author} {\bibfnamefont {N.~C.}\ \bibnamefont {Handy}},\ }\bibfield  {title} {\bibinfo {title} {On the determination of excitation energies using density functional theory},\ }\href {https://doi.org/10.1039/A910321J} {\bibfield  {journal} {\bibinfo  {journal} {Phys. Chem. Chem. Phys.}\ }\textbf {\bibinfo {volume} {2}},\ \bibinfo {pages} {2117} (\bibinfo {year} {2000})}\BibitemShut {NoStop}%
\bibitem [{\citenamefont {Chong}\ \emph {et~al.}(2002)\citenamefont {Chong}, \citenamefont {Gritsenko},\ and\ \citenamefont {Baerends}}]{Chong021760}%
  \BibitemOpen
  \bibfield  {author} {\bibinfo {author} {\bibfnamefont {D.~P.}\ \bibnamefont {Chong}}, \bibinfo {author} {\bibfnamefont {O.~V.}\ \bibnamefont {Gritsenko}},\ and\ \bibinfo {author} {\bibfnamefont {E.~J.}\ \bibnamefont {Baerends}},\ }\bibfield  {title} {\bibinfo {title} {Interpretation of the {Kohn}{\textendash}{Sham} orbital energies as approximate vertical ionization potentials},\ }\href {https://doi.org/10.1063/1.1430255} {\bibfield  {journal} {\bibinfo  {journal} {The Journal of Chemical Physics}\ }\textbf {\bibinfo {volume} {116}},\ \bibinfo {pages} {1760} (\bibinfo {year} {2002})}\BibitemShut {NoStop}%
\bibitem [{\citenamefont {Gritsenko}\ \emph {et~al.}(2003)\citenamefont {Gritsenko}, \citenamefont {Bra\"{\i}da},\ and\ \citenamefont {Baerends}}]{Gritsenko031937a}%
  \BibitemOpen
  \bibfield  {author} {\bibinfo {author} {\bibfnamefont {O.~V.}\ \bibnamefont {Gritsenko}}, \bibinfo {author} {\bibfnamefont {B.}~\bibnamefont {Bra\"{\i}da}},\ and\ \bibinfo {author} {\bibfnamefont {E.~J.}\ \bibnamefont {Baerends}},\ }\bibfield  {title} {\bibinfo {title} {Physical interpretation and evaluation of the {Kohn-Sham} and {Dyson} components of the $\varepsilon$-{I} relations between the {Kohn-Sham} orbital energies and the ionization potentials},\ }\href {https://doi.org/10.1063/1.1582839} {\bibfield  {journal} {\bibinfo  {journal} {The Journal of Chemical Physics}\ }\textbf {\bibinfo {volume} {119}},\ \bibinfo {pages} {1937} (\bibinfo {year} {2003})}\BibitemShut {NoStop}%
\bibitem [{\citenamefont {Savin}\ \emph {et~al.}(1998)\citenamefont {Savin}, \citenamefont {Umrigar},\ and\ \citenamefont {Gonze}}]{Savin98391}%
  \BibitemOpen
  \bibfield  {author} {\bibinfo {author} {\bibfnamefont {A.}~\bibnamefont {Savin}}, \bibinfo {author} {\bibfnamefont {C.~J.}\ \bibnamefont {Umrigar}},\ and\ \bibinfo {author} {\bibfnamefont {X.}~\bibnamefont {Gonze}},\ }\bibfield  {title} {\bibinfo {title} {Relationship of {Kohn}-{Sham} eigenvalues to excitation energies},\ }\href {https://doi.org/10.1016/S0009-2614(98)00316-9} {\bibfield  {journal} {\bibinfo  {journal} {Chemical Physics Letters}\ }\textbf {\bibinfo {volume} {288}},\ \bibinfo {pages} {391} (\bibinfo {year} {1998})}\BibitemShut {NoStop}%
\bibitem [{\citenamefont {van Meer}\ \emph {et~al.}(2014)\citenamefont {van Meer}, \citenamefont {Gritsenko},\ and\ \citenamefont {Baerends}}]{vanMeer144432}%
  \BibitemOpen
  \bibfield  {author} {\bibinfo {author} {\bibfnamefont {R.}~\bibnamefont {van Meer}}, \bibinfo {author} {\bibfnamefont {O.~V.}\ \bibnamefont {Gritsenko}},\ and\ \bibinfo {author} {\bibfnamefont {E.~J.}\ \bibnamefont {Baerends}},\ }\bibfield  {title} {\bibinfo {title} {Physical {Meaning} of {Virtual} {Kohn}{\textendash}{Sham} {Orbitals} and {Orbital} {Energies}: {An} {Ideal} {Basis} for the {Description} of {Molecular} {Excitations}},\ }\href {https://doi.org/10.1021/ct500727c} {\bibfield  {journal} {\bibinfo  {journal} {Journal of Chemical Theory and Computation}\ }\textbf {\bibinfo {volume} {10}},\ \bibinfo {pages} {4432} (\bibinfo {year} {2014})}\BibitemShut {NoStop}%
\bibitem [{\citenamefont {Cohen}\ \emph {et~al.}(2008{\natexlab{c}})\citenamefont {Cohen}, \citenamefont {Mori-S{\'a}nchez},\ and\ \citenamefont {Yang}}]{Cohen08115123}%
  \BibitemOpen
  \bibfield  {author} {\bibinfo {author} {\bibfnamefont {A.~J.}\ \bibnamefont {Cohen}}, \bibinfo {author} {\bibfnamefont {P.}~\bibnamefont {Mori-S{\'a}nchez}},\ and\ \bibinfo {author} {\bibfnamefont {W.}~\bibnamefont {Yang}},\ }\bibfield  {title} {\bibinfo {title} {Fractional charge perspective on the band gap in density-functional theory},\ }\href {https://doi.org/10/fp3sk9} {\bibfield  {journal} {\bibinfo  {journal} {Physical Review B}\ }\textbf {\bibinfo {volume} {77}},\ \bibinfo {pages} {115123} (\bibinfo {year} {2008}{\natexlab{c}})}\BibitemShut {NoStop}%
\bibitem [{\citenamefont {Yang}\ \emph {et~al.}(2012)\citenamefont {Yang}, \citenamefont {Cohen},\ and\ \citenamefont {Mori-S{\'a}nchez}}]{Yang12204111}%
  \BibitemOpen
  \bibfield  {author} {\bibinfo {author} {\bibfnamefont {W.}~\bibnamefont {Yang}}, \bibinfo {author} {\bibfnamefont {A.~J.}\ \bibnamefont {Cohen}},\ and\ \bibinfo {author} {\bibfnamefont {P.}~\bibnamefont {Mori-S{\'a}nchez}},\ }\bibfield  {title} {\bibinfo {title} {Derivative discontinuity, bandgap and lowest unoccupied molecular orbital in density functional theory},\ }\href {https://doi.org/10.1063/1.3702391} {\bibfield  {journal} {\bibinfo  {journal} {The Journal of Chemical Physics}\ }\textbf {\bibinfo {volume} {136}},\ \bibinfo {pages} {204111} (\bibinfo {year} {2012})}\BibitemShut {NoStop}%
\bibitem [{\citenamefont {Perdew}\ \emph {et~al.}(2017)\citenamefont {Perdew}, \citenamefont {Yang}, \citenamefont {Burke}, \citenamefont {Yang}, \citenamefont {Gross}, \citenamefont {Scheffler}, \citenamefont {Scuseria}, \citenamefont {Henderson}, \citenamefont {Zhang}, \citenamefont {Ruzsinszky}, \citenamefont {Peng}, \citenamefont {Sun}, \citenamefont {Trushin},\ and\ \citenamefont {Goerling}}]{Perdew172801}%
  \BibitemOpen
  \bibfield  {author} {\bibinfo {author} {\bibfnamefont {J.~P.}\ \bibnamefont {Perdew}}, \bibinfo {author} {\bibfnamefont {W.}~\bibnamefont {Yang}}, \bibinfo {author} {\bibfnamefont {K.}~\bibnamefont {Burke}}, \bibinfo {author} {\bibfnamefont {Z.}~\bibnamefont {Yang}}, \bibinfo {author} {\bibfnamefont {E.~K.~U.}\ \bibnamefont {Gross}}, \bibinfo {author} {\bibfnamefont {M.}~\bibnamefont {Scheffler}}, \bibinfo {author} {\bibfnamefont {G.~E.}\ \bibnamefont {Scuseria}}, \bibinfo {author} {\bibfnamefont {T.~M.}\ \bibnamefont {Henderson}}, \bibinfo {author} {\bibfnamefont {I.~Y.}\ \bibnamefont {Zhang}}, \bibinfo {author} {\bibfnamefont {A.}~\bibnamefont {Ruzsinszky}}, \bibinfo {author} {\bibfnamefont {H.}~\bibnamefont {Peng}}, \bibinfo {author} {\bibfnamefont {J.}~\bibnamefont {Sun}}, \bibinfo {author} {\bibfnamefont {E.}~\bibnamefont {Trushin}},\ and\ \bibinfo {author} {\bibfnamefont {A.}~\bibnamefont {Goerling}},\ }\bibfield  {title} {\bibinfo {title} {Understanding band gaps of solids in generalized
  {{Kohn-Sham}} theory},\ }\href {https://doi.org/10.1073/pnas.1621352114} {\bibfield  {journal} {\bibinfo  {journal} {Proc. Natl. Acad. Sci. U. S. A.}\ }\textbf {\bibinfo {volume} {114}},\ \bibinfo {pages} {2801} (\bibinfo {year} {2017})},\ \Eprint {https://arxiv.org/abs/28265085} {28265085} \BibitemShut {NoStop}%
\bibitem [{\citenamefont {Su}\ \emph {et~al.}(2020{\natexlab{b}})\citenamefont {Su}, \citenamefont {Mahler},\ and\ \citenamefont {Yang}}]{Su2020}%
  \BibitemOpen
  \bibfield  {author} {\bibinfo {author} {\bibfnamefont {N.~Q.}\ \bibnamefont {Su}}, \bibinfo {author} {\bibfnamefont {A.}~\bibnamefont {Mahler}},\ and\ \bibinfo {author} {\bibfnamefont {W.}~\bibnamefont {Yang}},\ }\bibfield  {title} {\bibinfo {title} {Preserving symmetry and degeneracy in the localized orbital scaling correction approach},\ }\href {https://doi.org/10.1021/acs.jpclett.9b03888} {\bibfield  {journal} {\bibinfo  {journal} {J. Phys. Chem. Lett.}\ }\textbf {\bibinfo {volume} {11}},\ \bibinfo {pages} {1528} (\bibinfo {year} {2020}{\natexlab{b}})}\BibitemShut {NoStop}%
\bibitem [{\citenamefont {Hirao}\ \emph {et~al.}(2021)\citenamefont {Hirao}, \citenamefont {Bae}, \citenamefont {Song},\ and\ \citenamefont {Chan}}]{Hirao213489}%
  \BibitemOpen
  \bibfield  {author} {\bibinfo {author} {\bibfnamefont {K.}~\bibnamefont {Hirao}}, \bibinfo {author} {\bibfnamefont {H.-S.}\ \bibnamefont {Bae}}, \bibinfo {author} {\bibfnamefont {J.-W.}\ \bibnamefont {Song}},\ and\ \bibinfo {author} {\bibfnamefont {B.}~\bibnamefont {Chan}},\ }\bibfield  {title} {\bibinfo {title} {Koopmans{\textquoteright}-{Type} {Theorem} in {Kohn}{\textendash}{Sham} {Theory} with {Optimally} {Tuned} {Long}-{Range}-{Corrected} ({LC}) {Functionals}},\ }\href {https://doi.org/10/gmvbtt} {\bibfield  {journal} {\bibinfo  {journal} {The Journal of Physical Chemistry A}\ }\textbf {\bibinfo {volume} {125}},\ \bibinfo {pages} {3489} (\bibinfo {year} {2021})}\BibitemShut {NoStop}%
\bibitem [{\citenamefont {Borghi}\ \emph {et~al.}(2014)\citenamefont {Borghi}, \citenamefont {Ferretti}, \citenamefont {Nguyen}, \citenamefont {Dabo},\ and\ \citenamefont {Marzari}}]{Borghi14075135}%
  \BibitemOpen
  \bibfield  {author} {\bibinfo {author} {\bibfnamefont {G.}~\bibnamefont {Borghi}}, \bibinfo {author} {\bibfnamefont {A.}~\bibnamefont {Ferretti}}, \bibinfo {author} {\bibfnamefont {N.~L.}\ \bibnamefont {Nguyen}}, \bibinfo {author} {\bibfnamefont {I.}~\bibnamefont {Dabo}},\ and\ \bibinfo {author} {\bibfnamefont {N.}~\bibnamefont {Marzari}},\ }\bibfield  {title} {\bibinfo {title} {Koopmans-compliant functionals and their performance against reference molecular data},\ }\href {https://doi.org/10.1103/PhysRevB.90.075135} {\bibfield  {journal} {\bibinfo  {journal} {Physical Review B}\ }\textbf {\bibinfo {volume} {90}},\ \bibinfo {pages} {075135} (\bibinfo {year} {2014})}\BibitemShut {NoStop}%
\bibitem [{\citenamefont {Colonna}\ \emph {et~al.}(2018)\citenamefont {Colonna}, \citenamefont {Nguyen}, \citenamefont {Ferretti},\ and\ \citenamefont {Marzari}}]{Colonna182549}%
  \BibitemOpen
  \bibfield  {author} {\bibinfo {author} {\bibfnamefont {N.}~\bibnamefont {Colonna}}, \bibinfo {author} {\bibfnamefont {N.~L.}\ \bibnamefont {Nguyen}}, \bibinfo {author} {\bibfnamefont {A.}~\bibnamefont {Ferretti}},\ and\ \bibinfo {author} {\bibfnamefont {N.}~\bibnamefont {Marzari}},\ }\bibfield  {title} {\bibinfo {title} {Screening in {Orbital}-{Density}-{Dependent} {Functionals}},\ }\href {https://doi.org/10/gdmtmg} {\bibfield  {journal} {\bibinfo  {journal} {Journal of Chemical Theory and Computation}\ }\textbf {\bibinfo {volume} {14}},\ \bibinfo {pages} {2549} (\bibinfo {year} {2018})}\BibitemShut {NoStop}%
\bibitem [{\citenamefont {Nguyen}\ \emph {et~al.}(2018)\citenamefont {Nguyen}, \citenamefont {Colonna}, \citenamefont {Ferretti},\ and\ \citenamefont {Marzari}}]{Nguyen18021051}%
  \BibitemOpen
  \bibfield  {author} {\bibinfo {author} {\bibfnamefont {N.~L.}\ \bibnamefont {Nguyen}}, \bibinfo {author} {\bibfnamefont {N.}~\bibnamefont {Colonna}}, \bibinfo {author} {\bibfnamefont {A.}~\bibnamefont {Ferretti}},\ and\ \bibinfo {author} {\bibfnamefont {N.}~\bibnamefont {Marzari}},\ }\bibfield  {title} {\bibinfo {title} {Koopmans-{Compliant} {Spectral} {Functionals} for {Extended} {Systems}},\ }\href {https://doi.org/10/gdr85n} {\bibfield  {journal} {\bibinfo  {journal} {Physical Review X}\ }\textbf {\bibinfo {volume} {8}},\ \bibinfo {pages} {021051} (\bibinfo {year} {2018})}\BibitemShut {NoStop}%
\bibitem [{\citenamefont {Colonna}\ \emph {et~al.}(2019)\citenamefont {Colonna}, \citenamefont {Nguyen}, \citenamefont {Ferretti},\ and\ \citenamefont {Marzari}}]{Colonna191905}%
  \BibitemOpen
  \bibfield  {author} {\bibinfo {author} {\bibfnamefont {N.}~\bibnamefont {Colonna}}, \bibinfo {author} {\bibfnamefont {N.~L.}\ \bibnamefont {Nguyen}}, \bibinfo {author} {\bibfnamefont {A.}~\bibnamefont {Ferretti}},\ and\ \bibinfo {author} {\bibfnamefont {N.}~\bibnamefont {Marzari}},\ }\bibfield  {title} {\bibinfo {title} {Koopmans-{Compliant} {Functionals} and {Potentials} and {Their} {Application} to the {GW100} {Test} {Set}},\ }\href {https://doi.org/10/gfvfvz} {\bibfield  {journal} {\bibinfo  {journal} {Journal of Chemical Theory and Computation}\ }\textbf {\bibinfo {volume} {15}},\ \bibinfo {pages} {1905} (\bibinfo {year} {2019})}\BibitemShut {NoStop}%
\bibitem [{\citenamefont {Colonna}\ \emph {et~al.}(2022{\natexlab{b}})\citenamefont {Colonna}, \citenamefont {De~Gennaro}, \citenamefont {Linscott},\ and\ \citenamefont {Marzari}}]{Colonna225435}%
  \BibitemOpen
  \bibfield  {author} {\bibinfo {author} {\bibfnamefont {N.}~\bibnamefont {Colonna}}, \bibinfo {author} {\bibfnamefont {R.}~\bibnamefont {De~Gennaro}}, \bibinfo {author} {\bibfnamefont {E.}~\bibnamefont {Linscott}},\ and\ \bibinfo {author} {\bibfnamefont {N.}~\bibnamefont {Marzari}},\ }\bibfield  {title} {\bibinfo {title} {Koopmans {Spectral} {Functionals} in {Periodic} {Boundary} {Conditions}},\ }\href {https://doi.org/10.1021/acs.jctc.2c00161} {\bibfield  {journal} {\bibinfo  {journal} {Journal of Chemical Theory and Computation}\ }\textbf {\bibinfo {volume} {18}},\ \bibinfo {pages} {5435} (\bibinfo {year} {2022}{\natexlab{b}})}\BibitemShut {NoStop}%
\bibitem [{\citenamefont {Williams}\ and\ \citenamefont {Yang}(2024)}]{Williams24}%
  \BibitemOpen
  \bibfield  {author} {\bibinfo {author} {\bibfnamefont {J.~Z.}\ \bibnamefont {Williams}}\ and\ \bibinfo {author} {\bibfnamefont {W.}~\bibnamefont {Yang}},\ }\href {https://doi.org/10.48550/arXiv.2406.07351} {\bibinfo {title} {Correcting {Delocalization} {Error} in {Materials} with {Localized} {Orbitals} and {Linear}-{Response} {Screening}}} (\bibinfo {year} {2024}),\ \bibinfo {note} {arXiv:2406.07351 [cond-mat, physics:physics]}\BibitemShut {NoStop}%
\bibitem [{\citenamefont {Yu}\ \emph {et~al.}(2024)\citenamefont {Yu}, \citenamefont {Mei}, \citenamefont {Chen},\ and\ \citenamefont {Yang}}]{Yu24}%
  \BibitemOpen
  \bibfield  {author} {\bibinfo {author} {\bibfnamefont {J.}~\bibnamefont {Yu}}, \bibinfo {author} {\bibfnamefont {Y.}~\bibnamefont {Mei}}, \bibinfo {author} {\bibfnamefont {Z.}~\bibnamefont {Chen}},\ and\ \bibinfo {author} {\bibfnamefont {W.}~\bibnamefont {Yang}},\ }\href {https://doi.org/10.48550/arXiv.2406.06345} {\bibinfo {title} {Accurate {Prediction} of {Core} {Level} {Binding} {Energies} from {Ground}-{State} {Density} {Functional} {Calculations}: {The} {Importance} of {Localization} and {Screening}}} (\bibinfo {year} {2024}),\ \bibinfo {note} {arXiv:2406.06345 [physics]}\BibitemShut {NoStop}%
\bibitem [{\citenamefont {Mei}\ \emph {et~al.}(2019)\citenamefont {Mei}, \citenamefont {Li}, \citenamefont {Su},\ and\ \citenamefont {Yang}}]{Mei19666}%
  \BibitemOpen
  \bibfield  {author} {\bibinfo {author} {\bibfnamefont {Y.}~\bibnamefont {Mei}}, \bibinfo {author} {\bibfnamefont {C.}~\bibnamefont {Li}}, \bibinfo {author} {\bibfnamefont {N.~Q.}\ \bibnamefont {Su}},\ and\ \bibinfo {author} {\bibfnamefont {W.}~\bibnamefont {Yang}},\ }\bibfield  {title} {\bibinfo {title} {Approximating {Quasiparticle} and {Excitation} {Energies} from {Ground} {State} {Generalized} {Kohn}{\textendash}{Sham} {Calculations}},\ }\href {https://doi.org/10/gfskbb} {\bibfield  {journal} {\bibinfo  {journal} {The Journal of Physical Chemistry A}\ }\textbf {\bibinfo {volume} {123}},\ \bibinfo {pages} {666} (\bibinfo {year} {2019})}\BibitemShut {NoStop}%
\bibitem [{\citenamefont {Koopmans}(1934)}]{Koopmans34104}%
  \BibitemOpen
  \bibfield  {author} {\bibinfo {author} {\bibfnamefont {T.}~\bibnamefont {Koopmans}},\ }\bibfield  {title} {\bibinfo {title} {Über die zuordnung von wellenfunktionen und eigenwerten zu den einzelnen elektronen eines atoms},\ }\href {https://doi.org/https://doi.org/10.1016/S0031-8914(34)90011-2} {\bibfield  {journal} {\bibinfo  {journal} {Physica}\ }\textbf {\bibinfo {volume} {1}},\ \bibinfo {pages} {104} (\bibinfo {year} {1934})}\BibitemShut {NoStop}%
\bibitem [{\citenamefont {Haiduke}\ and\ \citenamefont {Bartlett}(2018)}]{Haiduke18131101}%
  \BibitemOpen
  \bibfield  {author} {\bibinfo {author} {\bibfnamefont {R.~L.~A.}\ \bibnamefont {Haiduke}}\ and\ \bibinfo {author} {\bibfnamefont {R.~J.}\ \bibnamefont {Bartlett}},\ }\bibfield  {title} {\bibinfo {title} {Communication: {Can} excitation energies be obtained from orbital energies in a correlated orbital theory?},\ }\href {https://doi.org/10/gfc4zf} {\bibfield  {journal} {\bibinfo  {journal} {The Journal of Chemical Physics}\ }\textbf {\bibinfo {volume} {149}},\ \bibinfo {pages} {131101} (\bibinfo {year} {2018})}\BibitemShut {NoStop}%
\bibitem [{\citenamefont {Mei}\ \emph {et~al.}(2018)\citenamefont {Mei}, \citenamefont {Li}, \citenamefont {Su},\ and\ \citenamefont {Yang}}]{Mei18b}%
  \BibitemOpen
  \bibfield  {author} {\bibinfo {author} {\bibfnamefont {Y.}~\bibnamefont {Mei}}, \bibinfo {author} {\bibfnamefont {C.}~\bibnamefont {Li}}, \bibinfo {author} {\bibfnamefont {N.~Q.}\ \bibnamefont {Su}},\ and\ \bibinfo {author} {\bibfnamefont {W.}~\bibnamefont {Yang}},\ }\bibfield  {title} {\bibinfo {title} {Approximating {Quasiparticle} and {Excitation} {Energies} from {Ground} {State} {Generalized} {Kohn}-{Sham} {Calculations}},\ }\href {http://arxiv.org/abs/1810.09906} {\bibfield  {journal} {\bibinfo  {journal} {arXiv:1810.09906 [physics]}\ } (\bibinfo {year} {2018})},\ \bibinfo {note} {arXiv: 1810.09906}\BibitemShut {NoStop}%
\bibitem [{\citenamefont {Mei}\ and\ \citenamefont {Yang}(2019{\natexlab{a}})}]{Mei19144109}%
  \BibitemOpen
  \bibfield  {author} {\bibinfo {author} {\bibfnamefont {Y.}~\bibnamefont {Mei}}\ and\ \bibinfo {author} {\bibfnamefont {W.}~\bibnamefont {Yang}},\ }\bibfield  {title} {\bibinfo {title} {Charge transfer excitation energies from ground state density functional theory calculations},\ }\href {https://doi.org/10/gfx8dk} {\bibfield  {journal} {\bibinfo  {journal} {The Journal of Chemical Physics}\ }\textbf {\bibinfo {volume} {150}},\ \bibinfo {pages} {144109} (\bibinfo {year} {2019}{\natexlab{a}})}\BibitemShut {NoStop}%
\bibitem [{\citenamefont {Mei}\ and\ \citenamefont {Yang}(2019{\natexlab{b}})}]{Mei192538}%
  \BibitemOpen
  \bibfield  {author} {\bibinfo {author} {\bibfnamefont {Y.}~\bibnamefont {Mei}}\ and\ \bibinfo {author} {\bibfnamefont {W.}~\bibnamefont {Yang}},\ }\bibfield  {title} {\bibinfo {title} {Excited-{State} {Potential} {Energy} {Surfaces}, {Conical} {Intersections}, and {Analytical} {Gradients} from {Ground}-{State} {Density} {Functional} {Theory}},\ }\href {https://doi.org/10/gf2jtz} {\bibfield  {journal} {\bibinfo  {journal} {The Journal of Physical Chemistry Letters}\ ,\ \bibinfo {pages} {2538}} (\bibinfo {year} {2019}{\natexlab{b}})}\BibitemShut {NoStop}%
\bibitem [{\citenamefont {Kuan}\ \emph {et~al.}(2024)\citenamefont {Kuan}, \citenamefont {Yeh}, \citenamefont {Yang},\ and\ \citenamefont {Hsu}}]{Kuan246126}%
  \BibitemOpen
  \bibfield  {author} {\bibinfo {author} {\bibfnamefont {K.-Y.}\ \bibnamefont {Kuan}}, \bibinfo {author} {\bibfnamefont {S.-H.}\ \bibnamefont {Yeh}}, \bibinfo {author} {\bibfnamefont {W.}~\bibnamefont {Yang}},\ and\ \bibinfo {author} {\bibfnamefont {C.-P.}\ \bibnamefont {Hsu}},\ }\bibfield  {title} {\bibinfo {title} {Excited-{State} {Charge} {Transfer} {Coupling} from {Quasiparticle} {Energy} {Density} {Functional} {Theory}},\ }\href {https://doi.org/10.1021/acs.jpclett.4c00850} {\bibfield  {journal} {\bibinfo  {journal} {The Journal of Physical Chemistry Letters}\ }\textbf {\bibinfo {volume} {15}},\ \bibinfo {pages} {6126} (\bibinfo {year} {2024})}\BibitemShut {NoStop}%
\bibitem [{\citenamefont {Yang}\ and\ \citenamefont {Ayers}(2024)}]{Yang24}%
  \BibitemOpen
  \bibfield  {author} {\bibinfo {author} {\bibfnamefont {W.}~\bibnamefont {Yang}}\ and\ \bibinfo {author} {\bibfnamefont {P.~W.}\ \bibnamefont {Ayers}},\ }\href {https://doi.org/10.48550/arXiv.2403.04604} {\bibinfo {title} {Foundation for the {$\Delta$}{SCF} {Approach} in {Density} {Functional} {Theory}}} (\bibinfo {year} {2024}),\ \bibinfo {note} {arXiv:2403.04604}\BibitemShut {NoStop}%
\bibitem [{\citenamefont {Yang}\ and\ \citenamefont {Fan}(2024)}]{yang_fractional_2024}%
  \BibitemOpen
  \bibfield  {author} {\bibinfo {author} {\bibfnamefont {W.}~\bibnamefont {Yang}}\ and\ \bibinfo {author} {\bibfnamefont {Y.}~\bibnamefont {Fan}},\ }\href {https://doi.org/10.48550/arXiv.2408.08443} {\bibinfo {title} {Fractional {Charges}, {Linear} {Conditions} and {Chemical} {Potentials} for {Excited} {States} in {$\Delta$}{SCF} {Theory}}} (\bibinfo {year} {2024}),\ \bibinfo {note} {arXiv:2408.08443}\BibitemShut {NoStop}%
\bibitem [{\citenamefont {Slater}(1972)}]{Slater1974AdvQuantumChem}%
  \BibitemOpen
  \bibfield  {author} {\bibinfo {author} {\bibfnamefont {J.~C.}\ \bibnamefont {Slater}},\ }\bibfield  {title} {\bibinfo {title} {Statistical exchange-correlation in the self-consistent field},\ }in\ \href {https://doi.org/https://doi.org/10.1016/S0065-3276(08)60541-9} {\emph {\bibinfo {booktitle} {Advances in quantum chemistry}}},\ Vol.~\bibinfo {volume} {6},\ \bibinfo {editor} {edited by\ \bibinfo {editor} {\bibfnamefont {P.-O.}\ \bibnamefont {L{\"o}wdin}}}\ (\bibinfo  {publisher} {Academic Press},\ \bibinfo {year} {1972})\ pp.\ \bibinfo {pages} {1--92}\BibitemShut {NoStop}%
\bibitem [{\citenamefont {Slater}\ and\ \citenamefont {Wood}(1970)}]{SlaterWood1970SlaterTM}%
  \BibitemOpen
  \bibfield  {author} {\bibinfo {author} {\bibfnamefont {J.~C.}\ \bibnamefont {Slater}}\ and\ \bibinfo {author} {\bibfnamefont {J.~H.}\ \bibnamefont {Wood}},\ }\bibfield  {title} {\bibinfo {title} {Statistical exchange and the total energy of a crystal},\ }\href {https://doi.org/https://doi.org/10.1002/qua.560050703} {\bibfield  {journal} {\bibinfo  {journal} {International Journal of Quantum Chemistry}\ }\textbf {\bibinfo {volume} {5}},\ \bibinfo {pages} {3} (\bibinfo {year} {1970})}\BibitemShut {NoStop}%
\bibitem [{\citenamefont {Ziegler}\ \emph {et~al.}(1977)\citenamefont {Ziegler}, \citenamefont {Rauk},\ and\ \citenamefont {Baerends}}]{ZieglerRaukBaerends1977deltaSCF}%
  \BibitemOpen
  \bibfield  {author} {\bibinfo {author} {\bibfnamefont {T.}~\bibnamefont {Ziegler}}, \bibinfo {author} {\bibfnamefont {A.}~\bibnamefont {Rauk}},\ and\ \bibinfo {author} {\bibfnamefont {E.~J.}\ \bibnamefont {Baerends}},\ }\bibfield  {title} {\bibinfo {title} {On the calculation of multiplet energies by the hartree-fock-slater method},\ }\href {https://doi.org/10.1007/BF00551551} {\bibfield  {journal} {\bibinfo  {journal} {Theoretica chimica acta}\ }\textbf {\bibinfo {volume} {43}},\ \bibinfo {pages} {261} (\bibinfo {year} {1977})}\BibitemShut {NoStop}%
\bibitem [{\citenamefont {Gavnholt}\ \emph {et~al.}(2008)\citenamefont {Gavnholt}, \citenamefont {Olsen}, \citenamefont {Engelund},\ and\ \citenamefont {Schi{\o}tz}}]{GavnholtOlsen2008DeltaSCFes}%
  \BibitemOpen
  \bibfield  {author} {\bibinfo {author} {\bibfnamefont {J.}~\bibnamefont {Gavnholt}}, \bibinfo {author} {\bibfnamefont {T.}~\bibnamefont {Olsen}}, \bibinfo {author} {\bibfnamefont {M.}~\bibnamefont {Engelund}},\ and\ \bibinfo {author} {\bibfnamefont {J.}~\bibnamefont {Schi{\o}tz}},\ }\bibfield  {title} {\bibinfo {title} {$\delta$ self-consistent field method to obtain potential energy surfaces of excited molecules on surfaces},\ }\href {https://doi.org/10.1103/PhysRevB.78.075441} {\bibfield  {journal} {\bibinfo  {journal} {Physical Review B}\ }\textbf {\bibinfo {volume} {78}},\ \bibinfo {pages} {075441} (\bibinfo {year} {2008})}\BibitemShut {NoStop}%
\bibitem [{\citenamefont {Gilbert}\ \emph {et~al.}(2008)\citenamefont {Gilbert}, \citenamefont {Besley},\ and\ \citenamefont {Gill}}]{GilbertBesleyGill2008MOM}%
  \BibitemOpen
  \bibfield  {author} {\bibinfo {author} {\bibfnamefont {A.~T.~B.}\ \bibnamefont {Gilbert}}, \bibinfo {author} {\bibfnamefont {N.~A.}\ \bibnamefont {Besley}},\ and\ \bibinfo {author} {\bibfnamefont {P.~M.~W.}\ \bibnamefont {Gill}},\ }\bibfield  {title} {\bibinfo {title} {Self-consistent field calculations of excited states using the maximum overlap method ({MOM})},\ }\href {https://doi.org/10.1021/jp801738f} {\bibfield  {journal} {\bibinfo  {journal} {J. Phys. Chem. A}\ }\textbf {\bibinfo {volume} {112}},\ \bibinfo {pages} {13164} (\bibinfo {year} {2008})}\BibitemShut {NoStop}%
\bibitem [{\citenamefont {Besley}\ \emph {et~al.}(2009)\citenamefont {Besley}, \citenamefont {Gilbert},\ and\ \citenamefont {Gill}}]{BesleyGilbertGill2009DeltaSCF}%
  \BibitemOpen
  \bibfield  {author} {\bibinfo {author} {\bibfnamefont {N.~A.}\ \bibnamefont {Besley}}, \bibinfo {author} {\bibfnamefont {A.~T.~B.}\ \bibnamefont {Gilbert}},\ and\ \bibinfo {author} {\bibfnamefont {P.~M.~W.}\ \bibnamefont {Gill}},\ }\bibfield  {title} {\bibinfo {title} {Self-consistent-field calculations of core excited states},\ }\href {https://doi.org/10.1063/1.3092928} {\bibfield  {journal} {\bibinfo  {journal} {The Journal of Chemical Physics}\ }\textbf {\bibinfo {volume} {130}},\ \bibinfo {pages} {124308} (\bibinfo {year} {2009})}\BibitemShut {NoStop}%
\bibitem [{\citenamefont {Harbola}\ and\ \citenamefont {Samal}(2009)}]{HarbolaSamal2009esHK}%
  \BibitemOpen
  \bibfield  {author} {\bibinfo {author} {\bibfnamefont {M.~K.}\ \bibnamefont {Harbola}}\ and\ \bibinfo {author} {\bibfnamefont {P.}~\bibnamefont {Samal}},\ }\bibfield  {title} {\bibinfo {title} {Time-independent excited-state density functional theory: study of 1s(2)2p(3)({S}-4) and 1s(2)2p(3)({D}-2) states of the boron isoelectronic series up to {Ne5}+},\ }\href@noop {} {\bibfield  {journal} {\bibinfo  {journal} {Journal of Physics B-Atomic Molecular and Optical Physics}\ }\textbf {\bibinfo {volume} {42}},\ \bibinfo {pages} {015003} (\bibinfo {year} {2009})}\BibitemShut {NoStop}%
\bibitem [{\citenamefont {Kowalczyk}\ \emph {et~al.}(2011)\citenamefont {Kowalczyk}, \citenamefont {Yost},\ and\ \citenamefont {Voorhis}}]{KowalczykanVoorhis2011deltaSCF}%
  \BibitemOpen
  \bibfield  {author} {\bibinfo {author} {\bibfnamefont {T.}~\bibnamefont {Kowalczyk}}, \bibinfo {author} {\bibfnamefont {S.~R.}\ \bibnamefont {Yost}},\ and\ \bibinfo {author} {\bibfnamefont {T.~V.}\ \bibnamefont {Voorhis}},\ }\bibfield  {title} {\bibinfo {title} {Assessment of the $\delta${SCF} density functional theory approach for electronic excitations in organic dyes},\ }\href {https://doi.org/10.1063/1.3530801} {\bibfield  {journal} {\bibinfo  {journal} {The Journal of Chemical Physics}\ }\textbf {\bibinfo {volume} {134}},\ \bibinfo {pages} {054128} (\bibinfo {year} {2011})}\BibitemShut {NoStop}%
\bibitem [{\citenamefont {Maurer}\ and\ \citenamefont {Reuter}(2011)}]{MaurerReuter2011deltaSCF}%
  \BibitemOpen
  \bibfield  {author} {\bibinfo {author} {\bibfnamefont {R.~J.}\ \bibnamefont {Maurer}}\ and\ \bibinfo {author} {\bibfnamefont {K.}~\bibnamefont {Reuter}},\ }\bibfield  {title} {\bibinfo {title} {Assessing computationally efficient isomerization dynamics: $\delta${SCF} density-functional theory study of azobenzene molecular switching},\ }\href {https://doi.org/10.1063/1.3664305} {\bibfield  {journal} {\bibinfo  {journal} {The Journal of Chemical Physics}\ }\textbf {\bibinfo {volume} {135}},\ \bibinfo {pages} {224303} (\bibinfo {year} {2011})}\BibitemShut {NoStop}%
\bibitem [{\citenamefont {Maurer}\ and\ \citenamefont {Reuter}(2013)}]{MaurerReuter2013deltaSCF}%
  \BibitemOpen
  \bibfield  {author} {\bibinfo {author} {\bibfnamefont {R.~J.}\ \bibnamefont {Maurer}}\ and\ \bibinfo {author} {\bibfnamefont {K.}~\bibnamefont {Reuter}},\ }\bibfield  {title} {\bibinfo {title} {Excited-state potential-energy surfaces of metal-adsorbed organic molecules from linear expansion $\delta$-self-consistent field density-functional theory ($\delta${SCF}-{DFT})},\ }\href {https://doi.org/10.1063/1.4812398} {\bibfield  {journal} {\bibinfo  {journal} {The Journal of Chemical Physics}\ }\textbf {\bibinfo {volume} {139}},\ \bibinfo {pages} {014708} (\bibinfo {year} {2013})}\BibitemShut {NoStop}%
\bibitem [{\citenamefont {Seidu}\ \emph {et~al.}(2015)\citenamefont {Seidu}, \citenamefont {Krykunov},\ and\ \citenamefont {Ziegler}}]{SeiduZiegler2015constrictedDFT}%
  \BibitemOpen
  \bibfield  {author} {\bibinfo {author} {\bibfnamefont {I.}~\bibnamefont {Seidu}}, \bibinfo {author} {\bibfnamefont {M.}~\bibnamefont {Krykunov}},\ and\ \bibinfo {author} {\bibfnamefont {T.}~\bibnamefont {Ziegler}},\ }\bibfield  {title} {\bibinfo {title} {Applications of time-dependent and time-independent density functional theory to rydberg transitions},\ }\href {https://doi.org/10.1021/jp5082802} {\bibfield  {journal} {\bibinfo  {journal} {The Journal of Physical Chemistry A}\ }\textbf {\bibinfo {volume} {119}},\ \bibinfo {pages} {5107} (\bibinfo {year} {2015})}\BibitemShut {NoStop}%
\bibitem [{\citenamefont {Ye}\ \emph {et~al.}(2017)\citenamefont {Ye}, \citenamefont {Welborn}, \citenamefont {Ricke},\ and\ \citenamefont {Van~Voorhis}}]{YeWelbornVanVoorhis2017deltaSCFalgorithm}%
  \BibitemOpen
  \bibfield  {author} {\bibinfo {author} {\bibfnamefont {H.-Z.}\ \bibnamefont {Ye}}, \bibinfo {author} {\bibfnamefont {M.}~\bibnamefont {Welborn}}, \bibinfo {author} {\bibfnamefont {N.~D.}\ \bibnamefont {Ricke}},\ and\ \bibinfo {author} {\bibfnamefont {T.}~\bibnamefont {Van~Voorhis}},\ }\bibfield  {title} {\bibinfo {title} {$\sigma$-{SCF}: {A} direct energy-targeting method to mean-field excited states},\ }\href {https://doi.org/10.1063/1.5001262} {\bibfield  {journal} {\bibinfo  {journal} {The Journal of Chemical Physics}\ }\textbf {\bibinfo {volume} {147}},\ \bibinfo {pages} {214104} (\bibinfo {year} {2017})}\BibitemShut {NoStop}%
\bibitem [{\citenamefont {Hait}\ and\ \citenamefont {Head-Gordon}(2020)}]{HaitHeadgordon2020deltaSCF}%
  \BibitemOpen
  \bibfield  {author} {\bibinfo {author} {\bibfnamefont {D.}~\bibnamefont {Hait}}\ and\ \bibinfo {author} {\bibfnamefont {M.}~\bibnamefont {Head-Gordon}},\ }\bibfield  {title} {\bibinfo {title} {Highly accurate prediction of core spectra of molecules at density functional theory cost: {Attaining} sub-electronvolt error from a restricted open-shell {Kohn}{\textendash}{Sham} approach},\ }\href {https://doi.org/10.1021/acs.jpclett.9b03661} {\bibfield  {journal} {\bibinfo  {journal} {The Journal of Physical Chemistry Letters}\ }\textbf {\bibinfo {volume} {11}},\ \bibinfo {pages} {775} (\bibinfo {year} {2020})}\BibitemShut {NoStop}%
\bibitem [{\citenamefont {Carter-Fenk}\ and\ \citenamefont {Herbert}(2020)}]{CarterfenkHerbert2020deltaSCFnumerical}%
  \BibitemOpen
  \bibfield  {author} {\bibinfo {author} {\bibfnamefont {K.}~\bibnamefont {Carter-Fenk}}\ and\ \bibinfo {author} {\bibfnamefont {J.~M.}\ \bibnamefont {Herbert}},\ }\bibfield  {title} {\bibinfo {title} {State-targeted energy projection: {A} simple and robust approach to orbital relaxation of non-aufbau self-consistent field solutions},\ }\href {https://doi.org/10.1021/acs.jctc.0c00502} {\bibfield  {journal} {\bibinfo  {journal} {Journal of Chemical Theory and Computation}\ }\textbf {\bibinfo {volume} {16}},\ \bibinfo {pages} {5067} (\bibinfo {year} {2020})}\BibitemShut {NoStop}%
\bibitem [{\citenamefont {Levi}\ \emph {et~al.}(2020)\citenamefont {Levi}, \citenamefont {Ivanov},\ and\ \citenamefont {J{\'o}nsson}}]{LeviJonsson2020deltaSCF}%
  \BibitemOpen
  \bibfield  {author} {\bibinfo {author} {\bibfnamefont {G.}~\bibnamefont {Levi}}, \bibinfo {author} {\bibfnamefont {A.~V.}\ \bibnamefont {Ivanov}},\ and\ \bibinfo {author} {\bibfnamefont {H.}~\bibnamefont {J{\'o}nsson}},\ }\bibfield  {title} {\bibinfo {title} {Variational calculations of excited states via direct optimization of the orbitals in {DFT}},\ }\href {https://doi.org/10.1039/D0FD00064G} {\bibfield  {journal} {\bibinfo  {journal} {Faraday Discussions}\ }\textbf {\bibinfo {volume} {224}},\ \bibinfo {pages} {448} (\bibinfo {year} {2020})}\BibitemShut {NoStop}%
\bibitem [{\citenamefont {Corzo}\ \emph {et~al.}(2022)\citenamefont {Corzo}, \citenamefont {Abou~Taka}, \citenamefont {Pribram-Jones},\ and\ \citenamefont {Hratchian}}]{CorzoPribramjonesHratchian2022PMOM}%
  \BibitemOpen
  \bibfield  {author} {\bibinfo {author} {\bibfnamefont {H.~H.}\ \bibnamefont {Corzo}}, \bibinfo {author} {\bibfnamefont {A.}~\bibnamefont {Abou~Taka}}, \bibinfo {author} {\bibfnamefont {A.}~\bibnamefont {Pribram-Jones}},\ and\ \bibinfo {author} {\bibfnamefont {H.~P.}\ \bibnamefont {Hratchian}},\ }\bibfield  {title} {\bibinfo {title} {Using projection operators with maximum overlap methods to simplify challenging self-consistent field optimization},\ }\href {https://doi.org/https://doi.org/10.1002/jcc.26797} {\bibfield  {journal} {\bibinfo  {journal} {Journal of Computational Chemistry}\ }\textbf {\bibinfo {volume} {43}},\ \bibinfo {pages} {382} (\bibinfo {year} {2022})}\BibitemShut {NoStop}%
\bibitem [{\citenamefont {Kumar}\ and\ \citenamefont {Luber}(2022)}]{KumarLuber2022deltaSCF}%
  \BibitemOpen
  \bibfield  {author} {\bibinfo {author} {\bibfnamefont {C.}~\bibnamefont {Kumar}}\ and\ \bibinfo {author} {\bibfnamefont {S.}~\bibnamefont {Luber}},\ }\bibfield  {title} {\bibinfo {title} {Robust $\delta${SCF} calculations with direct energy functional minimization methods and {STEP} for molecules and materials},\ }\href {https://doi.org/10.1063/5.0075927} {\bibfield  {journal} {\bibinfo  {journal} {The Journal of Chemical Physics}\ }\textbf {\bibinfo {volume} {156}},\ \bibinfo {pages} {154104} (\bibinfo {year} {2022})}\BibitemShut {NoStop}%
\bibitem [{\citenamefont {Vandaele}\ \emph {et~al.}(2022)\citenamefont {Vandaele}, \citenamefont {Mali{\v s}},\ and\ \citenamefont {Luber}}]{VandaeleLuber2002deltaSCFreview}%
  \BibitemOpen
  \bibfield  {author} {\bibinfo {author} {\bibfnamefont {E.}~\bibnamefont {Vandaele}}, \bibinfo {author} {\bibfnamefont {M.}~\bibnamefont {Mali{\v s}}},\ and\ \bibinfo {author} {\bibfnamefont {S.}~\bibnamefont {Luber}},\ }\bibfield  {title} {\bibinfo {title} {The $\delta${SCF} method for non-adiabatic dynamics of systems in the liquid phase},\ }\href {https://doi.org/10.1063/5.0083340} {\bibfield  {journal} {\bibinfo  {journal} {The Journal of Chemical Physics}\ }\textbf {\bibinfo {volume} {156}},\ \bibinfo {pages} {130901} (\bibinfo {year} {2022})}\BibitemShut {NoStop}%
\bibitem [{\citenamefont {Parr}\ \emph {et~al.}(1978)\citenamefont {Parr}, \citenamefont {Donnelly}, \citenamefont {Levy},\ and\ \citenamefont {Palke}}]{Parr783801}%
  \BibitemOpen
  \bibfield  {author} {\bibinfo {author} {\bibfnamefont {R.~G.}\ \bibnamefont {Parr}}, \bibinfo {author} {\bibfnamefont {R.~A.}\ \bibnamefont {Donnelly}}, \bibinfo {author} {\bibfnamefont {M.}~\bibnamefont {Levy}},\ and\ \bibinfo {author} {\bibfnamefont {W.~E.}\ \bibnamefont {Palke}},\ }\bibfield  {title} {\bibinfo {title} {Electronegativity: the density functional viewpoint},\ }\href@noop {} {\bibfield  {journal} {\bibinfo  {journal} {The Journal of Chemical Physics}\ }\textbf {\bibinfo {volume} {68}},\ \bibinfo {pages} {3801} (\bibinfo {year} {1978})}\BibitemShut {NoStop}%
\bibitem [{\citenamefont {Mori-S{\'a}nchez}\ \emph {et~al.}(2006)\citenamefont {Mori-S{\'a}nchez}, \citenamefont {Cohen},\ and\ \citenamefont {Yang}}]{Mori-Sanchez06201102}%
  \BibitemOpen
  \bibfield  {author} {\bibinfo {author} {\bibfnamefont {P.}~\bibnamefont {Mori-S{\'a}nchez}}, \bibinfo {author} {\bibfnamefont {A.~J.}\ \bibnamefont {Cohen}},\ and\ \bibinfo {author} {\bibfnamefont {W.}~\bibnamefont {Yang}},\ }\bibfield  {title} {\bibinfo {title} {Many-electron self-interaction error in approximate density functionals},\ }\href {https://doi.org/10.1063/1.2403848} {\bibfield  {journal} {\bibinfo  {journal} {The Journal of Chemical Physics}\ }\textbf {\bibinfo {volume} {125}},\ \bibinfo {pages} {201102} (\bibinfo {year} {2006})}\BibitemShut {NoStop}%
\bibitem [{\citenamefont {Ruzsinszky}\ \emph {et~al.}(2006)\citenamefont {Ruzsinszky}, \citenamefont {Perdew}, \citenamefont {Csonka}, \citenamefont {Vydrov},\ and\ \citenamefont {Scuseria}}]{Ruzsinszky06194112}%
  \BibitemOpen
  \bibfield  {author} {\bibinfo {author} {\bibfnamefont {A.}~\bibnamefont {Ruzsinszky}}, \bibinfo {author} {\bibfnamefont {J.~P.}\ \bibnamefont {Perdew}}, \bibinfo {author} {\bibfnamefont {G.~I.}\ \bibnamefont {Csonka}}, \bibinfo {author} {\bibfnamefont {O.~A.}\ \bibnamefont {Vydrov}},\ and\ \bibinfo {author} {\bibfnamefont {G.~E.}\ \bibnamefont {Scuseria}},\ }\bibfield  {title} {\bibinfo {title} {Spurious fractional charge on dissociated atoms: {Pervasive} and resilient self-interaction error of common density functionals},\ }\href {https://doi.org/10.1063/1.2387954} {\bibfield  {journal} {\bibinfo  {journal} {Journal of Chemical Physics}\ }\textbf {\bibinfo {volume} {125}},\ \bibinfo {pages} {194112} (\bibinfo {year} {2006})}\BibitemShut {NoStop}%
\bibitem [{\citenamefont {Becke}(1988)}]{Becke1988}%
  \BibitemOpen
  \bibfield  {author} {\bibinfo {author} {\bibfnamefont {A.~D.}\ \bibnamefont {Becke}},\ }\bibfield  {title} {\bibinfo {title} {Density-functional exchange-energy approximation with correct asymptotic behavior},\ }\href {https://doi.org/10.1103/PhysRevA.38.3098} {\bibfield  {journal} {\bibinfo  {journal} {Phys. Rev. A}\ }\textbf {\bibinfo {volume} {38}},\ \bibinfo {pages} {3098} (\bibinfo {year} {1988})}\BibitemShut {NoStop}%
\bibitem [{\citenamefont {Lee}\ \emph {et~al.}(1988)\citenamefont {Lee}, \citenamefont {Yang},\ and\ \citenamefont {Parr}}]{Lee1988}%
  \BibitemOpen
  \bibfield  {author} {\bibinfo {author} {\bibfnamefont {C.}~\bibnamefont {Lee}}, \bibinfo {author} {\bibfnamefont {W.}~\bibnamefont {Yang}},\ and\ \bibinfo {author} {\bibfnamefont {R.~G.}\ \bibnamefont {Parr}},\ }\bibfield  {title} {\bibinfo {title} {Development of the colle-salvetti correlation-energy formula into a functional of the electron density},\ }\href {https://doi.org/10.1103/PhysRevB.37.785} {\bibfield  {journal} {\bibinfo  {journal} {Phys. Rev. B}\ }\textbf {\bibinfo {volume} {37}},\ \bibinfo {pages} {785} (\bibinfo {year} {1988})}\BibitemShut {NoStop}%
\bibitem [{\citenamefont {Mei}\ \emph {et~al.}(2021)\citenamefont {Mei}, \citenamefont {Chen},\ and\ \citenamefont {Yang}}]{Mei2021}%
  \BibitemOpen
  \bibfield  {author} {\bibinfo {author} {\bibfnamefont {Y.}~\bibnamefont {Mei}}, \bibinfo {author} {\bibfnamefont {Z.}~\bibnamefont {Chen}},\ and\ \bibinfo {author} {\bibfnamefont {W.}~\bibnamefont {Yang}},\ }\bibfield  {title} {\bibinfo {title} {Exact second-order corrections and accurate quasiparticle energy calculations in density functional theory},\ }\href {https://doi.org/10.1021/acs.jpclett.1c01962} {\bibfield  {journal} {\bibinfo  {journal} {J. Phys. Chem. Lett.}\ }\textbf {\bibinfo {volume} {12}},\ \bibinfo {pages} {7236} (\bibinfo {year} {2021})}\BibitemShut {NoStop}%
\end{thebibliography}%

\end{document}